\documentclass[prd,nofootinbib,twocolumn,a4paper,floatfix,preprintnumbers,showpacs]{revtex4}
\usepackage{graphicx}
\usepackage{amsfonts}
\usepackage{amssymb}
\usepackage{amsbsy}
\usepackage{amsmath}
\usepackage{latexsym}
\usepackage{bm}
\usepackage{subfigure}
\usepackage{hyperref}

\begin{document}

\preprint{hep-th/0602194}

\title{Gravitational waves and cosmological braneworlds: \\ a characteristic
evolution scheme}

\author{Sanjeev S.~Seahra}
\email{sanjeev.seahra@port.ac.uk} \affiliation{Institute of
Cosmology \& Gravitation, University of Portsmouth, Portsmouth, PO1
2EG, UK}

\setlength\arraycolsep{2pt}
\newcommand*{\di}{\partial}
\newcommand*{\ds}[1]{ds^2_\text{\tiny{($#1$)}}}
\newcommand*{\ads}[1]{{AdS$_{#1}$}}
\newcommand*{\rb}{r_\text{b}}
\newcommand*{\xb}{x_\text{b}}
\newcommand*{\DOT}[2]{{\bm{{#1} \cdot {#2}}}}
\newcommand*{\dlangle}{\langle\!\langle}
\newcommand*{\drangle}{\rangle\!\rangle}
\newcommand*{\scriminus}{{\mathcal{I}^-}}

\date{July 17, 2006}

\begin{abstract}

Motivated by the problem of the evolution of bulk gravitational
waves in Randall-Sundrum cosmology, we develop a characteristic
numerical scheme to solve 1+1 dimensional wave equations in the
presence of a moving timelike boundary.  The scheme exhibits
quadratic convergence, is capable of handling arbitrary brane
trajectories, and is easily extendible to non-AdS bulk geometries.
We use our method to contrast two different prescriptions for the
bulk fluctuation initial conditions found in the literature;
namely, those of Hiramatsu et al.~and Ichiki and Nakamura. We find
that if the initial data surface is set far enough in the past,
the late time waveform on the brane is insensitive to the choice
between the two possibilities; and we present numeric and analytic
evidence that this phenomenon generalizes to more generic initial
data. Observationally, the main consequence of this work is to
re-affirm previous claims that the stochastic gravitational wave
spectrum is predominantly flat $\Omega_\text{GW} \propto f^0$, in
contradiction with naive predictions from the effective
4-dimensional theory. Furthermore, this flat spectrum result is
predicted to be robust against uncertainties in (or modifications
of) the bulk initial data, provided that the energy scale of brane
inflation is high enough.

\end{abstract}


\maketitle

\defcitealias{Hiramatsu:2004aa}{HKT}
\defcitealias{Ichiki:2004sx}{IN}
\defcitealias{Kobayashi:2005dd}{KT}

\section{Introduction}

It is well known that the Randall-Sundrum (RS) braneworld model
\cite{Randall:1999ee,Randall:1999vf} is in excellent agreement
with general relativity at low energies. This is the principal
appeal of the model; it is one of the only examples of a scenario
involving a large extra dimension that entails no serious
conflicts with general relativity.  However, this means that one
needs to consider high energy or strong gravity scenarios to
properly test the model. One possibility is to examine the high
energy epoch of braneworld cosmology, where exact solutions of the
5-dimensional field equations are known.  Well understood
braneworld phenomena \cite[review]{Maartens:2003tw} include a
modified cosmic expansion and early times and `dark radiation'
effects, whereby the Weyl curvature of the bulk projected on the
brane acts as an additional geometric source in the Friedmann
equation.

But if one wants to move beyond the exact description of the
background geometry in these cosmological models, there are
significant technical difficulties.  A cosmological brane is
essentially a moving boundary in a static 5-dimensional background
--- anti-de Sitter space in the RS model --- so perturbations are
described by bulk wave equations with
boundary conditions enforced on a non-trivial timelike surface.
While it is possible to make some analytic progress when the brane
is moving `slowly'
\cite{Easther:2003re,Battye:2003ks,Kobayashi:2004wy,Battye:2004qw},
the more interesting case of a fast-moving, high-energy brane
remains impervious to such treatment.

The purpose of this paper is to present a new numeric algorithm to
solve wave equations in the presence of a moving boundary.  For
the sake of simplicity, we restrict ourselves to a class of wave
equations and boundary conditions that correspond to tensor, or
gravitational wave (GW), perturbations.  This is not the first
attempt to deal with these equations numerically: previous efforts
include pseudo-spectral
\cite{Hiramatsu:2003iz,Hiramatsu:2004aa,Hiramatsu:2006bd} and
direct evolution
\cite{Ichiki:2003hf,Ichiki:2004sx,Kobayashi:2005jx,Kobayashi:2005dd,Kobayashi:2006pe}
methods using various null and non-null coordinate systems in
which the brane is stationary.

So why introduce another method?  Our primary motivation is to
develop an algorithm that offers several improvements to the
preexisting efforts.  Our technique is based on the highly
successful characteristic integration methods from black hole
perturbation theory \cite{Winicour:2005ge}.  These offer several
advantages, not the least of which is an elegance of
implementation that takes the causal structure of the spacetime
explicitly into account.  We also work in Poincar\'e coordinates,
which greatly simplify the bulk wave equation and avoid the
coordinate singularities that plague Gaussian normal patches. This
makes our algorithm both transparent and unique: while other
groups carry out their analysis in Poincar\'e coordinates, they
always transform the brane to a static location for actual
calculations.  Our procedure is developed from first principles,
and we pay careful attention to discretization errors. Hence we
have a good theoretical understanding of the convergence
properties of our method, which can then be tested in actual
calculations. The fact that the code behaves as expected
--- with explicit quadratic convergence --- imparts a certain level of
confidence in conclusions drawn from our numerical results.
Finally, our techniques should be easily adaptable to other
braneworld situations; i.e., more complicated bulk geometries,
sophisticated specification of initial conditions, multiple
branes, etc.

Our secondary motivation stems from the fact that the numerical
results found in the literature do not seem to agree with one
another.  In this paper, we limit the discussion to tensor type
perturbations, so the principal observational consequence of our
work is the present-day spectral density $\Omega_\text{GW}$ of
relic GWs generated during inflation on the brane
\cite{Maartens:1999hf,Langlois:2000ns,Gen:2000nu,Gorbunov:2001ge,Kobayashi:2003cn,%
Koyama:2004ap,Koyama:2005ek}. This `stochastic GW background' is
potentially observable by next-generation detectors such as the
Big Bang Observatory. (\citet{Maggiore:1999vm} offers
comprehensive review of the stochastic GW background from a
4-dimensional perspective.) Hiramatsu et
al.~\cite{Hiramatsu:2004aa} (henceforth
\citetalias{Hiramatsu:2004aa}) have predicted that
$\Omega_\text{GW} \propto f^0$ for frequencies above some
threshold $f_c$, whose value is determined by the curvature scale
of the bulk.  On the other hand, Ichiki and Nakamura
\cite{Ichiki:2003hf,Ichiki:2004sx} (henceforth
\citetalias{Ichiki:2004sx}) have predicted $\Omega_\text{GW}
\propto f^{-0.46}$ from their simulations using a different
initial condition. Recently, \citet{Kobayashi:2005dd} (henceforth
\citetalias{Kobayashi:2005dd}) have applied a different numerical
method to the quantum mechanical version of the problem, where one
treats the entire evolution of tensor modes during inflation and
radiation-domination as a particle-production phenomenon.  They
also derive an approximately flat spectrum for high frequencies,
in agreement with \citetalias{Hiramatsu:2004aa}.

What is the right answer?  It is difficult to compare these
calculations directly because each group uses a different
prescription for dealing with initial conditions \emph{and} a
different numerical method.  It is sensible that when trying to
solve a problem numerically, one should confirm that several
independent methods yield the same results under the \emph{same}
circumstances.  Only then can we be confident in the predictions.
To this end, we attempt to reproduce the results of
\citetalias{Hiramatsu:2004aa} and \citetalias{Ichiki:2004sx} using
our numerical method and their respective choices of initial
conditions.\footnote{The Wronskian formulation favoured by
\citetalias{Kobayashi:2005dd} is sufficiently distinct from the
other methods to defer its consideration to future work.}  We find
that our numerics reproduce the \citetalias{Hiramatsu:2004aa}
results within an acceptable tolerance, but we are unable to
duplicate the GW spectrum predicted by \citetalias{Ichiki:2004sx}.
Indeed, we find that as long as simulations are started in
sufficiently high energy epochs, both the
\citetalias{Hiramatsu:2004aa} and \citetalias{Ichiki:2004sx}
initial conditions lead to the same flat spectrum.

This is an interesting and somewhat unexpected result:
Superficially, the \citetalias{Hiramatsu:2004aa},
\citetalias{Ichiki:2004sx}, and \citetalias{Kobayashi:2005dd}
formulations appear quite different from one another, yet they
ultimately produce the same late-time behaviour.  This leads to
the question: How robust is the prediction of a flat GW spectrum
to arbitrary modification of the initial conditions? Answering
this is the third motivation for this paper, and is a fairly
ambitious goal.  This is because `arbitrary modification' implies
the need to consider an infinite number of cases, which is highly
impractical.  Hence, we need to settle for qualitative conclusions
drawn from numeric simulations of individual cases coupled with
some approximate analytic results.

The layout of the paper is as follows:  In
Sec.~\ref{sec:statement}, we describe the problem we are going to
solve in both general terms, and as specialized to tensor
perturbations in RS cosmologies.  The numeric algorithm used
throughout the paper is developed in Sec.~\ref{sec:numeric}, and
then comprehensively tested in Sec.~\ref{sec:code tests}. The
issue of initial conditions for GWs in braneworld cosmology is
reviewed in Sec.~\ref{sec:initial conditions}, and the
\citetalias{Hiramatsu:2004aa} and \citetalias{Ichiki:2004sx}
approaches are explicitly contrasted in Sec.~\ref{sec:HKT vs IN}.
A more generic class of initial data is considered in
Sec.~\ref{sec:generic initial data} using a combination of numeric
and analytic methods.  Sec.~\ref{sec:conclusions} is reserved for
our conclusions.  Appendix \ref{sec:GW background} reviews the
jargon associated with the stochastic GW background and how to
convert the results of numeric simulations into observational
predictions.

We use units in which $\hbar = c = 1$ and the `mostly positive'
metric signature.

\section{Statement of the problem}\label{sec:statement}

\subsection{Generic formulation}\label{sec:generic}

\begin{figure}
\begin{center}
\includegraphics{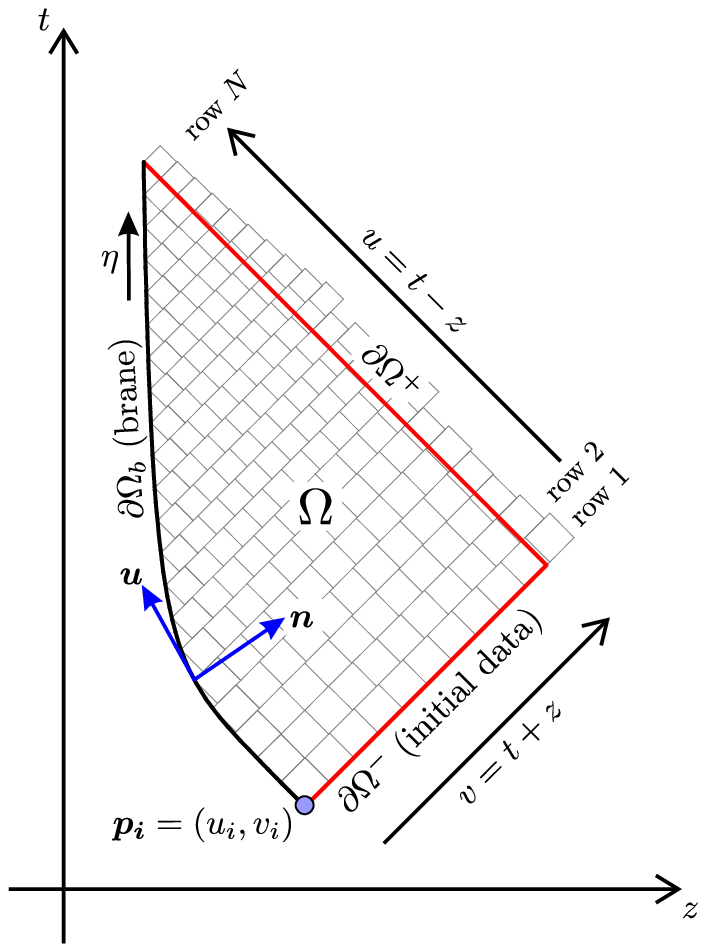}\\
\caption{(Colour online.) The spacetime domain $\Omega$ over which
we seek a numeric solution of the wave equation. Superimposed on
$\Omega$ is a (particularly coarse) example of the computational
grid we use to discretize the problem}\label{fig:grid}
\end{center}
\end{figure}

In this subsection, we define the generic type of problem that we
solve in this paper.  Consider the following wave equation:
\begin{equation}\label{eq:wave}
    [-\bm{D}^2 + V(z)]\psi(t,z) = 0.
\end{equation}
Here, $D_\alpha$ is a covariant derivative operator on a
\emph{flat} 2-dimensional manifold:
\begin{equation}
    ds^2 = -dt^2+dz^2 = - du\,dv,
\end{equation}
where $u = t - z$ and $v = t +z$ are the usual retarded and advanced
time coordinates.  Our goal is to obtain the value of the field
throughout a finite region $\Omega$ of the $(t,z)$ spacetime, which
is depicted in Fig.~\ref{fig:grid}.

As can be seen in this figure, the boundary of $\Omega$ is
composed of three distinct parts: $\di\Omega^\pm$ are null
surfaces to the future and past, while $\di\Omega_b$ is a timelike
surface that we will refer to as the `brane'.  The brane is
defined parametrically via the equations
\begin{equation}
    t = t_b(\eta), \quad z = z_b(\eta).
\end{equation}
The parameter $\eta$ is selected to be affine in the flat $(t,z)$
geometry:
\begin{equation}
    \DOT{u}{\di} = \dot t_b \, \di_t + \dot z_b \, \di_z, \quad \DOT{u}{u} = -1.
\end{equation}
Here and henceforth, we have use an overdot to denote $d/d\eta$.
As long as $z_b \ne$ constant, the brane can said to be `moving'
with respect to the $(t,z)$ frame of reference.  We also define a
normal vector to $\di\Omega_b$ pointing \emph{into} $\Omega$:
\begin{equation}\label{eq:normal defn}
    \DOT{u}{\di} = \dot z_b \, \di_t + \dot t_b \, \di_z, \quad \DOT{n}{n} = 1,
    \quad \DOT{u}{n} = 0.
\end{equation}

In order to have a well-defined hyperbolic problem, we must also
specify initial and boundary conditions.  The initial conditions
consist of choosing the value of the field $\psi$ on the past null
boundary $\di\Omega^-$.  This initial profile can be selected
arbitrarily.  We take the boundary conditions on the brane to be
\begin{equation}\label{eq:boundary condition}
    0 = [(\DOT{n}{D}) \, \psi - \alpha(\eta) \psi]_b.
\end{equation}
At this stage, we leave $\alpha$ as a finite, but otherwise
arbitrary, function of `time' on the brane.  Therefore, this
represents a wide variety of possible boundary conditions, except
for pure Dirichlet $\psi(t_b,z_b) = 0$.

\subsection{Application: gravitational waves in
brane cosmology}\label{sec:RS problem}

In this subsection, we show how the equations governing tensor
perturbations in RS one brane cosmologies can be written in the
general form introduced in Sec.~\ref{sec:generic} above.

A cosmological RS brane is identified as a hypersurface in
5-dimension anti-deSitter space with cosmological constant
$-6/\ell^2$:
\begin{equation}\label{eq:5D ads metric}
    ds^2 = \frac{\ell^2}{z^2} (-dt^2 + \delta_{ij} dx^i dx^j +
    dz^2).
\end{equation}
The brane is defined by the parametric equations
\begin{equation}
    t = t_b(\eta), \quad z = z_b(\eta),
\end{equation}
such that
\begin{equation}\label{eq:t_b dot}
    \dot t_b = \sqrt{1 + \dot z_b^2}.
\end{equation}
This equation ensures that the metric on the brane,
\begin{equation}
    ds^2 = a^2(\eta)(-d\eta^2 + \delta_{ij} dx^i dx^j),
\end{equation}
is of the standard cosmological form with $\eta$ as the conformal
time, and the scale factor identified as
\begin{equation}
    a(\eta) = \ell/z_b(\eta).
\end{equation}

The brane's dynamics are described by a Friedmann equation derived
from the Israel junction conditions:
\begin{equation}\label{eq:ordinary Friedmann}
    H^2(a) = \frac{\dot a^2}{a^4} = \frac{\dot z_b^2}{\ell^2} = \frac{\kappa_4^2}{3} \rho(a)
    \left[ 1 + \frac{\rho(a)}{2\lambda} \right].
\end{equation}
Here, $\rho$ is the density of brane matter, $\kappa_4^2 = 8\pi G_4$
is the 4-dimensional gravity-matter coupling, and $\lambda$ is the
brane tension.  We have enforced the RS fine tuning condition,
\begin{equation}
    \lambda \kappa_4^2 \ell^2 = 6,
\end{equation}
which means that there is no net cosmological constant on the
brane. We assume a single component perfect fluid for the brane
matter, with equation of state $\rho = w p$, which implies $\rho
\propto a^{-3(1+w)}$, as usual.  This allows us to write the
Friedmann equation as [$q \equiv 3(1+w)$]:
\begin{equation}\label{eq:Friedmann}
    (H\ell)^2 = \dot z_b^2 = \epsilon_* \left( \frac{z_b}{z_*}
    \right)^{q} \left[ 2 + \epsilon_* \left( \frac{z_b}{z_*}
    \right)^{q} \right].
\end{equation}
Here, $\epsilon_*$ is the energy density of brane matter,
normalized by the brane tension $\lambda$, at some reference epoch
$z_b = z_*$; i.e., $\epsilon_* = \rho_*/\lambda$.  Of course,
$z_*$ is freely a specifiable length scale.

Now, we turn our attention to tensor perturbations.  These are
defined by the substitution
\begin{equation}
    \delta_{ij} \rightarrow \delta_{ij} + \frac{1}{(2\pi M_5)^3} \!\! \sum_{\tiny{A=+,\times}}
    \! \int d^3k \, h(t,z;\mathbf{k},A) e^{i\mathbf{k}
    \cdot \mathbf{x}} \varepsilon^{(A)}_{ij},
\end{equation}
in the 5-dimensional metric (\ref{eq:5D ads metric}).  Here,
$\mathbf{k}$ is a 3-dimensional wavevector,
$\varepsilon^{(A)}_{ij} = \varepsilon^{(A)}_{ij}
(\mathbf{\hat{k}})$ is a constant transverse trace-free
polarization 3-tensor orthogonal to $\mathbf{k}$, and the
summation is over polarizations.  The 5-dimensional Planck mass
satisfies $\ell \kappa_4^2 M_5^3 = 1$.  Unless explicitly
required, we will omit the $\mathbf{k}$ and $A$ arguments from the
Fourier amplitude $h$ below. One finds that $h$ obeys
\begin{subequations}\label{eq:RS eqns}
\begin{eqnarray}
    \label{eq:RS wave} 0 & = & - \frac{\di^2 h}{\di t^2} + \frac{\di ^2 h}{\di z^2} - \frac{3}{z}
    \frac{\di h}{\di z} - k^2 h, \\
    \label{eq:RC BC} 0 & = & \left( H\ell \, \frac{\di h}{\di t} - \sqrt{1+H^2\ell^2}
    \frac{\di h}{\di z} \right)_b .
\end{eqnarray}
\end{subequations}
Now, to put these equations into the standard form of
Sec.~\ref{sec:generic}, we just need to make the
definition\footnote{For our numeric work, it is convenient to
characterize GWs by the $\psi$ wavefunction, as opposed to $h$.
But in the literature it has become standard to express
perturbations in terms of $h$, so we will always report our
results in as $h(t,z)$ instead of $\psi(t,z)$.  Of course, it is
trivial to move between the two descriptions using (\ref{eq:h psi
relation}).}
\begin{equation}\label{eq:h psi relation}
    \psi(t,z) = \left(\frac{z_*}{z}\right)^{3/2} h(t,z).
\end{equation}
Then, making use of $\dot z_b = -H\ell$ and eq.~(\ref{eq:t_b
dot}), we find that Eqs.~(\ref{eq:RS eqns}) reduce to
\begin{subequations}\label{eq:wave eqns}
\begin{eqnarray}
    0 & = & [-\bm{D}^2 + V(z)]\psi, \\
    0 & = & [(\DOT{n}{D}) \, \psi - \alpha(\eta) \psi]_b.
\end{eqnarray}
\end{subequations}
respectively, with
\begin{subequations}
\begin{eqnarray}
    V(z) & = & k^2 + \frac{15}{4z^2}, \\
    \alpha(\eta) & = & -\frac{3}{2}\frac{\dot{t}_b}{z_b} = -\frac{3}{2} \frac{\sqrt{1+H^2\ell^2}}{z_b}.
\end{eqnarray}
\end{subequations}
Here, $n^a$ is defined by (\ref{eq:normal defn}).  Hence, we have
successfully transformed the RS gravitational wave equation into an
equivalent form defined in the flat $(t,z)$ 2-manifold. We call
(\ref{eq:wave eqns}) the `canonical wave equation' governing RS
gravitational waves.

It is convenient to select $z_*$ to characterize the epoch when a
perturbation with a given wavenumber $k$ enters the horizon:
\begin{equation}\label{eq:k-epsilon connection}
    k = a_* H_*  \quad \Rightarrow \quad kz_* = H_* \ell =
    \sqrt{\epsilon_* (2 + \epsilon_*)}.
\end{equation}
If we then work with the dimensionless variables
\begin{equation}\label{eq:dimensionless}
    \hat t = t/z_*, \quad \tilde z = z/z_*, \quad \tilde\eta =
    \eta/z_*, \quad \tilde k = k z_*,
\end{equation}
the brane equations of motion, Eqs.~(\ref{eq:t_b dot}) and
(\ref{eq:Friedmann}), and gravitational wave equations,
Eq.~(\ref{eq:wave eqns}), are completely specified by the single
parameter $\epsilon_*$.  It is useful to define a critical value
$\epsilon_c = \sqrt{2} - 1 \approx 0.41$ that is associated with a
perturbation that enters the horizon when $H_* \ell = 1$.  We can
then classify modes as either long ($\epsilon_* < \epsilon_c$) or
short ($\epsilon_* > \epsilon_c$) wavelength when compared to the
characteristic length scale $\ell$ of the extra dimension.
Intuitively, we expect the 5-dimensional effects to be important
for short wavelength modes that enter the horizon when the
universe is smaller than $\ell$.

\begin{figure}
\begin{center}
\includegraphics{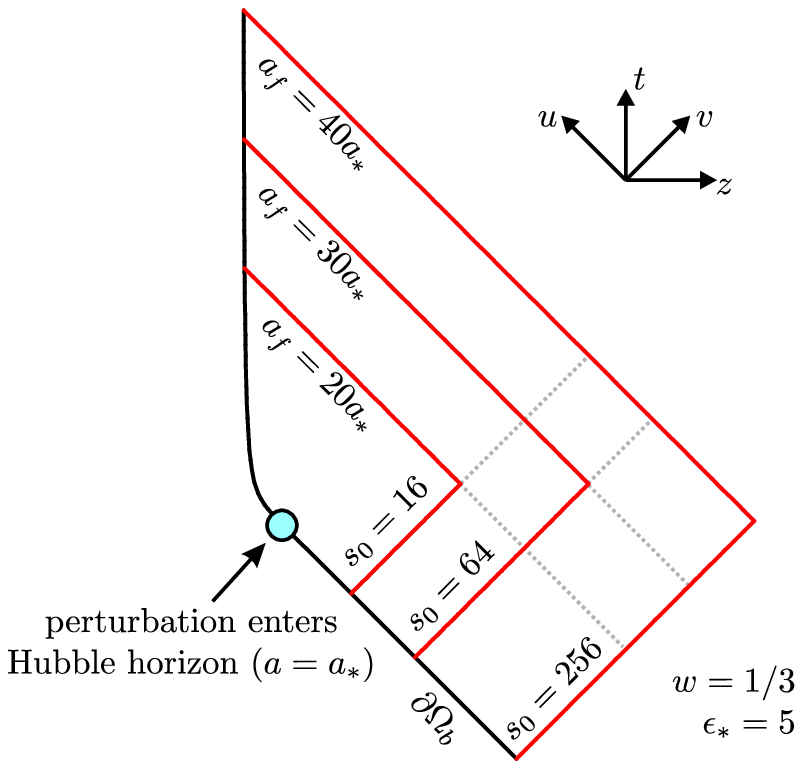}\\
\caption{(Colour online.) Illustration of how the parameters $s_0$
and $a_f/a_*$ move the past and future null boundaries of the
computational domain, respectively.  Here, we have taken a
radiation-dominated brane, so $\di\Omega^-$ is pushed further into
the past as $s_0$ is increased.  (The opposite is true for a
vacuum-dominated or deSitter brane, for example.)  All other
parameters being equal, the amount of CPU time required to complete
one simulation scales with the area of $\Omega$ divided by the
typical area of one of the cells shown in
Fig.~\ref{fig:grid}.}\label{fig:domain}
\end{center}
\end{figure}

Finally, to finishing specifying the problem, we need to fix the
position of the $\di\Omega^\pm$ boundaries of the computational
domain in Fig.~\ref{fig:grid}. This is equivalent to selecting the
initial and final brane size: $z_i = z_b(\eta_i)$ and $z_f =
z_b(\eta_f)$. It is useful to characterize the initial time by a
dimensionless number $s_0$, which is the ratio of the
perturbation's wavelength normalized by the horizon size at the
initial time:
\begin{equation}
    s_0 = \frac{a_i H_i}{k} = \frac{z_*}{z_i} \frac{H_i
    \ell}{H_*\ell}.
\end{equation}
Hence, by choosing $s_0$ and $\epsilon_*$, one determines $\tilde
z_i = z_i/z_*$.  One can place the future boundary of $\Omega$ by
selecting the ratio of the final brane size to the size at the
horizon-crossing time $a_f/a_* = 1/\tilde{z}_f$.  Therefore, in
dimensionless coordinates, the computational problem in
Sec.~\ref{sec:generic} is completely specified by
$(\epsilon_*,s_0,a_f/a_*)$, up to the choice of initial conditions
on $\di\Omega^-$.  In Fig.~\ref{fig:domain}, we show how the
choices of $s_0$ and $a_f/a_*$ alter the shape of the
computational domain for a radiation dominated brane ($w = 1/3$).

\section{Numeric scheme}\label{sec:numeric}

In this section, we develop a numerical scheme for solving the
class of 1+1 dimensional wave equations introduced in
Sec.~\ref{sec:generic}.  In subsequent sections, we will apply
this scheme to the RS tensor perturbation problem described in
Sec.~\ref{sec:RS problem}.

\subsection{Computational grid}

We begin by discretizing the computational domain $\Omega$ into a
finite number of `cells' as shown in Fig.~\ref{fig:grid}.  Each
cell is either a \emph{diamond} whose boundary consists of four
null segments of equal `size', or a \emph{triangle} whose boundary
is made up of two null and one timelike segment. The cells are
arranged into rows bounded by surfaces of constant $u$ such that
all diamonds in a given row have uniform size. Also, each row
contains one triangle where it intersects the brane at its
leftmost extreme. (In the example of Fig.~1, the triangles in the
early rows are very narrow and hard to see.)  The timelike segment
in each triangular cell gives a straight line approximation to the
brane trajectory.

Note that the diamond size generally varies from row to row. This is
because we have demanded that the $u =$ constant row boundaries
intersect the brane at evenly-spaced intervals of coordinate time
$t$.  The magnitude of this spacing is given by the parameter
$\Delta$; i.e., the future boundary of the $i^\text{th}$ row
intersects the brane at $t_i = t_0 + i\Delta$, where $t_0$ is some
initial time. Note that this is not the only possible choice; for
example, we could have demanded each row be regularly spaced in $u$,
$\eta$, or some other coordinate. This particular choice of spacing
ensures that each diamond cell has an area less than $2\Delta^2$,
while each triangular cell has an area less than $\Delta^2/4$.

\subsection{Diamond cellular evolution}

\begin{figure*}
\begin{center}
\subfigure[\,\,Diamond]{\includegraphics[scale=0.9]{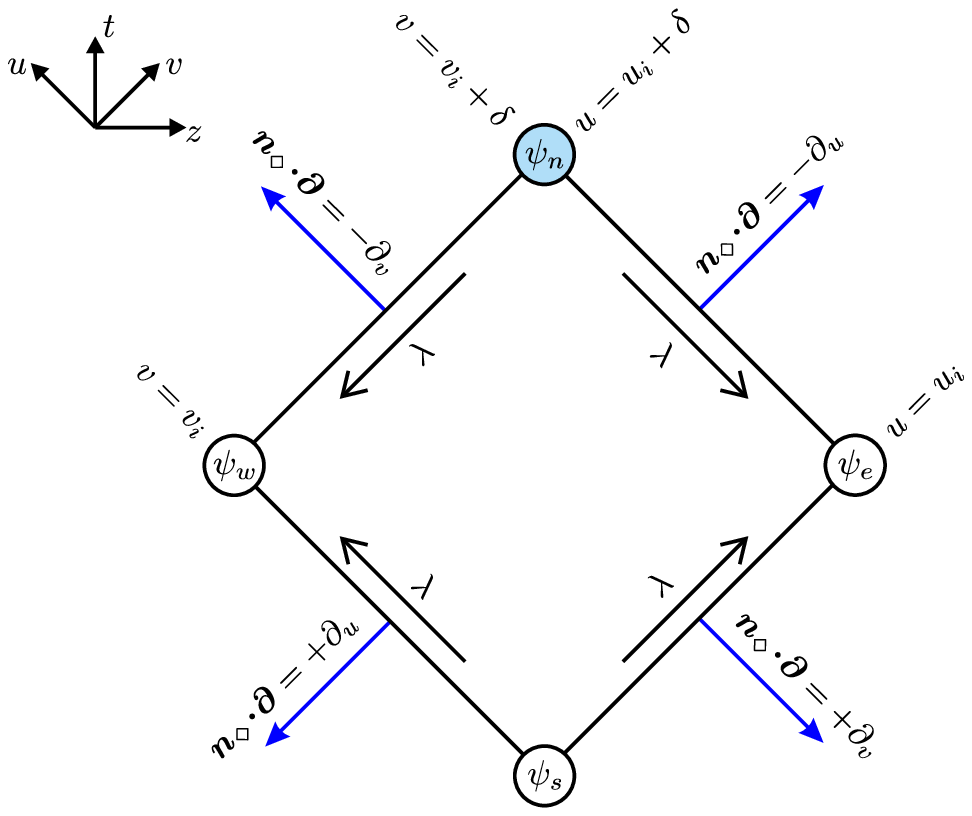}
\label{fig:diamond}} \qquad \qquad \subfigure[\,\,Timelike
triangle]{\includegraphics[scale=0.9]{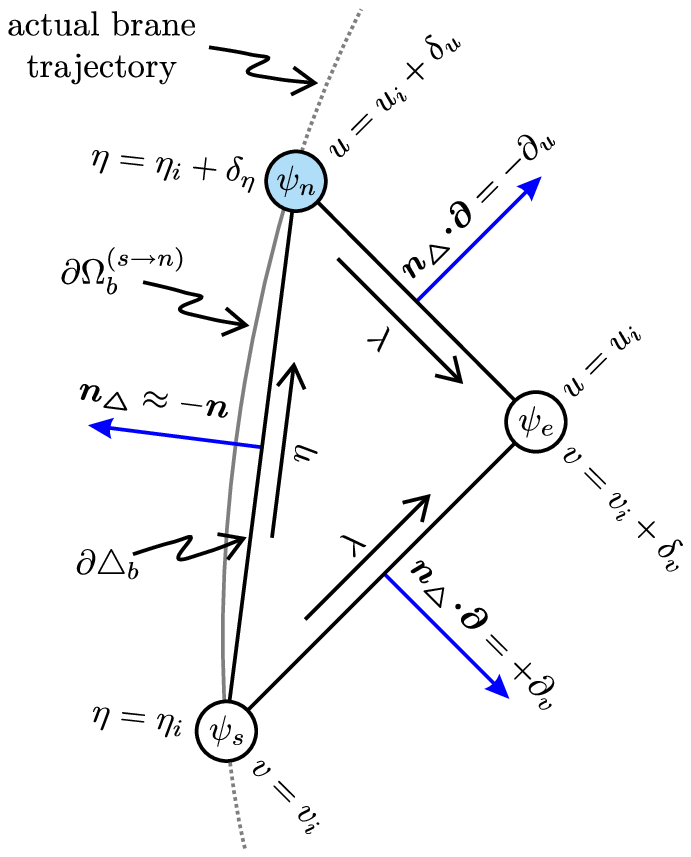}\label{fig:triangle}}
\caption{(Colour online.) Cellular geometries}
\end{center}
\end{figure*}

We now derive formulae that relate the value of the field at each
node in a particular cell---these will form the heart of the
integration scheme introduced in the next subsection.  First,
consider a typical diamond cell shown in Fig.~\ref{fig:diamond}.
This type of cell has four nodes that we label by their compass
directions: north, south, east, and west. The value of the field at
each node is denoted by $\psi_n$, $\psi_s$, $\psi_e$, and $\psi_w$,
respectively. We integrate the wave equation (\ref{eq:wave}) over
the cell and use Gauss's law to obtain:
\begin{equation}\label{eq:diamond integral}
    \int_{\di\Diamond} (\DOT{n_\Diamond}{D}) \, \psi = \int_{\Diamond} V \psi.
\end{equation}
Here, $\bm{n_\Diamond}$ is the \emph{outward} pointing normal to
the cell boundary $\di\Diamond$.  In each integral, the natural
`volume' element on the respective submanifolds is understood.
Because all of the boundary surfaces are null in a diamond cell,
$\bm{n_\Diamond}$ is actually everywhere tangent to $\di\Diamond$.
If $\lambda$ is an affine parameter along each segment of the
boundary, then we obtain
\begin{equation}
    n_\Diamond^\alpha = \frac{dx^\alpha}{d\lambda} \quad \Rightarrow \quad
    (\DOT{n_\Diamond}{D}) \, \psi = \frac{d\psi}{d\lambda}.
\end{equation}
This implies that the boundary term can be evaluated exactly by
integrating over the four null line segments composing
$\di\Diamond$ individually.  The result is:
\begin{equation}\label{eq:diamond boundary}
    \int_{\di\Diamond} (\DOT{n_\Diamond}{D}) \, \psi = 2(\psi_e + \psi_w -
    \psi_n - \psi_s).
\end{equation}
Note that when dealing with each segment, it is important to
integrate in the direction of increasing affine parameter
$\lambda$, which is indicated in Fig.~\ref{fig:diamond} by the
interior arrows.\footnote{Also note that this result could have
been obtained via the explicit double integral $\int_{\Diamond}
\bm{D}^2 \psi = \int \! \int du \, dv \,  (-2 \di_u \di_v \psi)$.
This is how this formula is usually derived for numeric problems
in black hole perturbation theory \cite{Lousto:1997wf}, but it is
difficult to generalize such an approach to triangular cells.}

By adopting a bilinear approximation for the integrand, the volume
integral in (\ref{eq:diamond integral}) is given by
\begin{equation}\label{eq:diamond volume}
    \int_{\Diamond} V \psi = \frac{\delta^2}{8} (V_n
    \psi_n + V_s \psi_s + V_e \psi_e + V_w \psi_w) +
    \mathcal{O}(\delta^4).
\end{equation}
Here, $V_n$ is the value of the potential at the northern node,
$V_s$ at the southern node, and so forth; hence, $V_n = V_s$.
Here, $\delta$ is the size of the cell in null coordinates.  Our
choice of grid spacing implies that
\begin{equation}
    \delta \le 2 \Delta \quad \Rightarrow \quad \delta =
    \mathcal{O}(\Delta).
\end{equation}
We now use (\ref{eq:diamond boundary}) and (\ref{eq:diamond
volume}) in (\ref{eq:diamond integral}) to isolate $\psi_n$:
\begin{equation}\label{eq:diamond evolution}
    \psi_n = -\psi_s + (\psi_w + \psi_e)(1-\tfrac{1}{8} \delta^2
    V_s) + \mathcal{O}(\Delta^4).
\end{equation}
Given the field value at the southern, eastern and western nodes,
this formula gives us the value at the northern node correct to
cubic order in $\Delta$.

Note that in order to derive (\ref{eq:diamond evolution}), we
performed a series expansion in $V_s\delta^2$ and retained the
first correction term.  Hence, one should only believe the
$\mathcal{O}(\Delta^4)$ error term in the diamond evolution law
when this approximation is valid; i.e., when
\begin{equation}\label{eq:potential condition}
    \delta^2 V(z) \lesssim \Delta^2 V(z) \ll 1.
\end{equation}
This is sensible requirement:  In order to achieve reliable
results, the characteristic size of a diamond $\Delta$ must be
much smaller then the characteristic length scale defined by the
potential $1/\sqrt{V(z)}$.

\subsection{Timelike triangle cellular evolution}

We now turn our attention to the timelike triangular cells at the
end of each row, an example of which is shown in
Fig.~\ref{fig:triangle}. This type of cell has three nodes: north,
south and east, and the boundary is composed of one timelike and two
null line segments. By construction, the brane's trajectory
precisely intersects the northern and southern nodes.  In the
calculation that follows, we view the timelike cell boundary
$\di\triangle_b$ as interchangeable with the actual brane trajectory
between these nodes, which we call $\di\Omega_b^{(s \rightarrow
n)}$.  Of course, there is some degree of error associated with such
an assumption, so we use a `$\approx$' sign to indicate equations
that are strictly true only when $\di\Omega_b^{(s \rightarrow n)} =
\di\triangle_b$.  Below, we discuss under which circumstances these
`$\approx$' signs can be regraded as equalities.

Integrating the wave equation over a triangular cell and applying
Gauss's law yields:
\begin{equation}\label{eq:triangle integral}
    \int_{\di\triangle} (\DOT{n_\triangle}{D}) \, \psi = \int_{\triangle} V \psi.
\end{equation}
As before, the null portions of the boundary integral can be
evaluated exactly.  For the timelike segment, we replace the path
$\di\triangle_b$ with $\di\Omega_b^{(s \rightarrow n)}$.  This
yields
\begin{equation}
    \int\limits_{\di\triangle} (\DOT{n_\triangle}{D}) \, \psi \approx 2 \psi_e - \psi_n - \psi_s -
    \!\!\!\!\!\! \int\limits_{\di\Omega_b^{(s \rightarrow n)}} \!\!\!\!\!\! (\DOT{n}{D}) \, \psi.
\end{equation}
Again, we label the field values by the node they are associated
with. Using the boundary condition (\ref{eq:boundary condition}),
we can substitute for the normal derivative in the brane integral.
Using the trapezoid approximation, we have
\begin{equation}
    \int\limits_{\di\Omega_b^{(s \rightarrow n)}} \!\!\!\!\!\!
    (\DOT{n}{D}) \, \psi = \frac{\delta_\eta}{2}(\alpha_n
    \psi_n + \alpha_s \psi_s) + \mathcal{O}(\delta_\eta^3).
\end{equation}
Here, $\alpha_n$ and $\alpha_s$ are the values of $\alpha(\eta)$ at
the northern and southern node, respectively.  The volume integral
over the cell is handled via a bilinear approximation as before:
\begin{equation}
    \int_{\triangle} V \psi = \frac{\delta_u \delta_v}{12} (V_n \psi_n + V_s \psi_s + V_e
    \psi_e) + \mathcal{O}[(\delta_u^2+\delta_v^2)^2].
\end{equation}
Now, let us denote coordinate differences between the northern and
southern nodes by $\delta_t$, $\delta_z$, etc.  Our choice of grid
spacing then gives
\begin{eqnarray}
    \nonumber \delta_t = \Delta & \Rightarrow & 0 \le \delta_\eta^2 \approx \delta_t^2 - \delta_z^2
    \le \Delta^2 \\ \nonumber & \Rightarrow & \delta_z \le \Delta \\
    \label{eq:triangle inequalities} & \Rightarrow &
    \delta_u^2 + \delta_v^2 \approx 2(\delta_\eta^2+2\delta_z^2) \le 6\Delta^2.
\end{eqnarray}
When we put Eqs.~(\ref{eq:triangle integral})--(\ref{eq:triangle
inequalities}) together, we obtain
\begin{eqnarray}\nonumber
    \psi_n & \approx & -{\frac {12+6\,\alpha_{{s}}\delta_{{\eta}}+\delta_{{u}}\delta_{{v}}V_{
{s}}}{12+6\,\alpha_{{n}}\delta_{{\eta}}+\delta_{{u}}\delta_{{v}}V_{{n}
}}} \psi_s \\ & & + {\frac {24-\delta_{{u}}\delta_{{v}}V_{{e}}
}{12+6\,\alpha_{{n}}
\delta_{{\eta}}+\delta_{{u}}\delta_{{v}}V_{{n}}}} \psi_e +
\mathcal{O}(\Delta^3) \label{eq:triangle evolution}.
\end{eqnarray}
Given field values at the southern and eastern nodes, this formula
gives $\psi_n$ accurate to quadratic order in $\Delta$, provided
that the discrepancy between $\di\Omega_b^{(s \rightarrow n)}$ and
$\di\triangle_b$ is negligible.

Under which circumstances can we ignore this discrepancy and
replace the above `$\approx$' signs with `$=$'?  Clearly, when
$\di\Omega_b^{(s \rightarrow n)}$ is well approximated by a
straight line throughout the cell.  In other words, when the
change $\bm{\delta u}$ in $\bm{u}$ over the cell is small. Note
that because the length of $\bm{u}$ is conserved, $\bm{\delta u}$
must be parallel to $\bm{n}$, so we really want $|\DOT{n}{\delta
u}| \ll 1$. This condition can be rewritten as
\begin{eqnarray}\nonumber
    |\DOT{n}{\delta u}| & \sim &  |n_\alpha u^\beta D_\beta u^\alpha | \delta_\eta
    \\ \nonumber & \lesssim & | u^\alpha u^\beta D_\beta n_\alpha| \Delta
    \\ & = & |\DOT{D}{n}| \Delta \ll 1.
\end{eqnarray}
This condition can be cast in a more geometric light by noting
that in 2 dimensions, the local radius of curvature of a curve is
given by $r_c = r_c(\eta) = 1/|\DOT{D}{n}|$.  Hence, the above
condition is equivalent to
\begin{equation}\label{eq:curvature condition}
    \Delta \ll r_c(\eta).
\end{equation}
In other words, we can reliably use the triangle evolution law
(\ref{eq:triangle evolution}) if the radius of curvature of the
brane is much larger than the characteristic size of the cell.

\subsection{The algorithm and theoretical
convergence}\label{sec:algorithm}

\begin{figure}
\begin{center}
\includegraphics[scale=0.9]{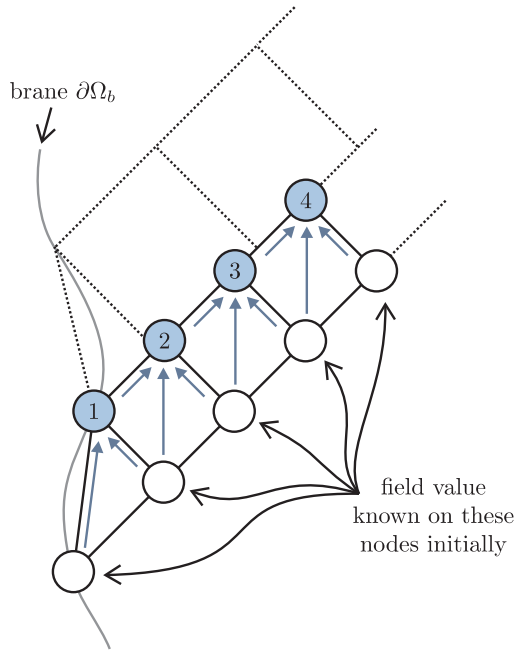}\\
\caption{(Colour online.) The algorithm used to deduce the field
value on the future boundary of a row, given the value on the past
boundary. The numbers indicate the order in which the nodal fields
are calculated, while the arrows inside the cells show how
information flows through the diagram; i.e., the field value at
nodes where arrows end depends directly on the field value at the
nodes where the arrows start}\label{fig:algorithm}
\end{center}
\end{figure}

Having defined our computational grid in Fig.~\ref{fig:grid} and
derived the diamond and triangular evolution laws,
(\ref{eq:diamond evolution}) and (\ref{eq:triangle evolution}),
our algorithm for solving the wave equation (\ref{eq:wave}) is
quite straightforward. Referring back to Fig.~\ref{fig:grid}, we
see that by specifying the value of $\psi$ on $\Omega^-$, we gain
knowledge of the field at the past boundary of row 1. An
enlargement of the situation is shown in Fig.~\ref{fig:algorithm}.
To obtain $\psi$ on the future boundary of the row, we first use
the triangle law to obtain the value at the node marked `1' in the
diagram. Then, we use the diamond evolution formula to fill in
node `2', then node `3', and so on.

After the field values on the future half of row 1 are found, we
then need to find the field at the nodes on the past boundary of the
next row. As can be seen in the diagram, those nodes do not line up
with the future nodes of row 1, so we must use an interpolation
scheme to determine the field there.  We use a polynomial
interpolation with a four-point stencil:  That is, for the
$i^\text{th}$ node on the past of row 2, we use the four closest
nodes on the future of row 1 to find a cubic polynomial
approximation to $\psi$.  That approximation is then used to
`fill-in' the value of $\psi$ at the node on the past boundary of
row 2.  This introduces an error of order $\mathcal{O}(\Delta^4)$ in
$\psi$.   Once this is accomplished for all the past nodes on row 2,
we then repeat the cycle by using the evolution laws to find the
field on the future of row 2, then interpolating to get $\psi$ on
the past of row 3, etc.

Since we have errors of order $\Delta^4$ in the evolution of bulk
cells and our interpolation from row to row, and errors of order
$\Delta^3$ in cells bordering the brane, we expect our overall
algorithm to exhibit quadratic convergence overall, provided that
the conditions (\ref{eq:potential condition}) and (\ref{eq:curvature
condition}) are met. That is, if the `exact' solution of the problem
is given by $\psi_\text{exact}$ while our numerical solution with
tolerance $\Delta$ is $\psi_\Delta$, we expect:
\begin{equation}\label{eq:convergence expectation}
    \psi_\Delta(t,z) - \psi_\text{exact}(t,z) = \Delta^2
    \varepsilon(t,z).
\end{equation}
Here, $\varepsilon(t,z)$ is a function that does not depend on
$\Delta$. We will test this convergence condition explicitly in
the next section.

\section{Code tests}\label{sec:code tests}

\subsection{Inertial (de Sitter) branes}\label{sec:de Sitter code
test}

In this subsection, we specialize to RS braneworld cosmological
models for which an exact solution to the tensor perturbation
problem of Sec.~\ref{sec:RS problem} is known.  Namely, we
consider de Sitter branes with $w = -1$.  Our goal is to compare
the results of the numerical scheme introduced in the previous
subsection to this exact solution, and thus test the reliability
of our algorithm.

\begin{figure}
\begin{center}
\includegraphics{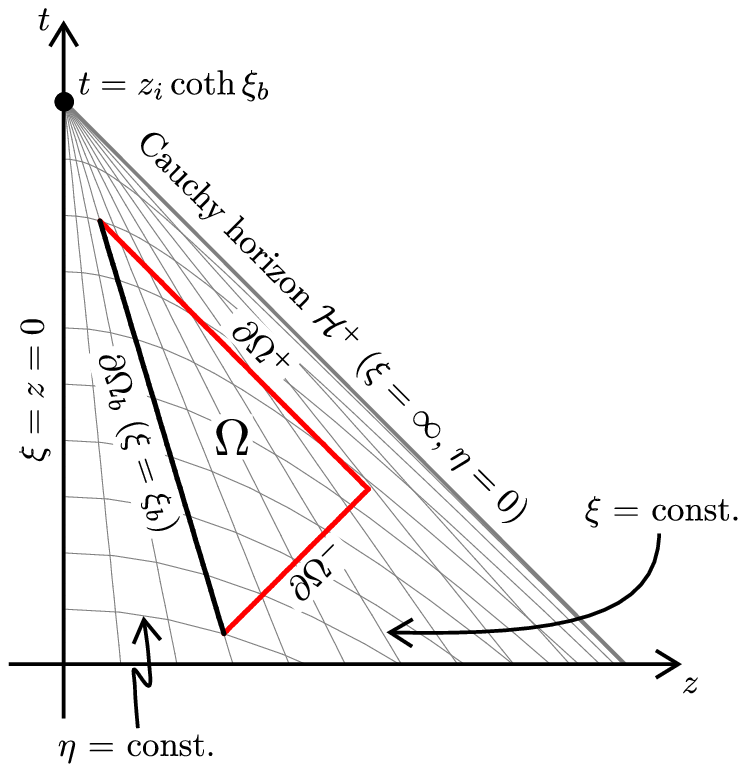}\\
\caption{(Colour online.) The level curves of the Rindler-like
$(\eta,\xi)$ coordinates in the $(t,z)$ plane.  A brane undergoing
pure-de Sitter inflation can be identified with any $\xi$ = constant
surface.  The solution to the tensor perturbation problem of
Sec.~\ref{sec:RS problem} is known exactly in these coordinates,
which allows us to test the accuracy and convergence of the numeric
algorithm developed in Sec.~\ref{sec:numeric}.  An example of the
computational domain $\Omega$ used for the numeric calculation is
also shown.}\label{fig:dS geometry}
\end{center}
\end{figure}

It is convenient to introduce a new coordinate system to describe
this scenario:
\begin{equation}
    t(\eta,\xi) = \eta \cosh \xi + z_i \coth \xi_b, \quad z(\eta,\xi) = -\eta \sinh \xi.
\end{equation}
Here, $\xi_b$ and $z_i$ are arbitrary positive constants, the
timelike coordinate $\eta$ is strictly negative, and the spacelike
coordinate satisfies $\xi \ge \xi_b$. Then, when $w=-1$ the brane
equations of motion (\ref{eq:t_b dot}) and (\ref{eq:Friedmann}) are
solved by the $\xi = \xi_b$ hypersurface; i.e.,
\begin{equation}
    t_b(\eta) = t(\eta,\xi_b), \quad z_b(\eta) =
    z(\eta,\xi_b).
\end{equation}
The Hubble constant on the brane is given by $H\ell = \sinh\xi_b$
and we find that the brane's speed,
\begin{equation}
    \frac{dz_b}{dt_b} = -\tanh\xi_b,
\end{equation}
is constant; i.e., the brane trajectory in the $(t,z)$ reference
frame is a straight line.  One could easily call such branes
`inertial'.  Also note that when $z_b = z_i$, we have $t_b = 0$.
In Fig.~\ref{fig:dS geometry}, we show the coordinate lines of the
$(\eta,\xi)$ patch in the $(t,z)$ plane.  The observant reader
will note that they are identical to the coordinate lines of
Rindler coordinates in Minkowski space; i.e., the $\xi =$ constant
lines represent a family of inertial observers whose trajectories
are converging to a point and the $\eta =$ constant curves
represent the surfaces over which their clocks are synchronized.

The advantage of introducing the $(\eta,\xi)$ coordinates is that
they render the wave equation and boundary condition (\ref{eq:RS
eqns}) separable. Hence, it is relatively straightforward to obtain
an exact solution for $h = h(\eta,\xi)$.  In general, this is given
by a superposition of mode functions $\{\phi_0,\phi_\nu\}$, with
$\nu \ge 0$.  The simplest of these is the so-called `zero-mode',
\begin{equation}\label{eq:de Sitter zero mode}
    \phi_0(\eta,\xi) = \frac{ C(\sinh\xi_b)\sinh\xi_b}{\ell \sqrt{2k\ell}} \left(\eta
    - \frac{i}{k} \right) e^{-ik\eta},
\end{equation}
where
\begin{equation}
    C(x) \equiv \left[ \sqrt{1+x^2} + x^2 \ln \left(
    \frac{x}{1+\sqrt{1+x^2}} \right) \right]^{-1/2}.
\end{equation}
Since $h = \phi_0$ is a legitimate solution of (\ref{eq:RS eqns}),
then
\begin{equation}
    \psi_\text{exact}(t,z) = \frac{z_*^{3/2}}{z^{3/2}} \text{Re}\,
    \left\{ \left[ k\eta(t, z) - i \right] e^{-i k \eta(t,z)}
    \right\}
\end{equation}
is a solution of the canonical wave equation (\ref{eq:wave eqns}),
with $\eta$ defined by
\begin{equation}
    \eta(t,z) = -\sqrt{(t-z_i \coth \xi_b)^2 - z^2}.
\end{equation}
We also define $h_\text{exact} = (z/z_*)^{3/2}\psi_\text{exact}$.

To test the algorithm, we proceed as follows: We set the
computational domain $\Omega$ (shown in Fig.~\ref{fig:dS
geometry}) by fixing a de Sitter brane trajectory, and then
selecting an initial and final time. In practice, we use the
dimensionless coordinates (\ref{eq:dimensionless}), so $\Omega$
and the dimensionless wavenumber $kz_*$ are determined by choosing
$(\epsilon_\star,s_0,a_f/a_*)$.  Then, our gird is defined by the
selection of a spacing $\Delta$. We synchronize our numeric
solution $\psi_\Delta$ to the exact solution on the initial time
hypersurface:
\begin{equation}
    \psi_\Delta(\di\Omega^-) = \psi_\text{exact}(\di\Omega^-).
\end{equation}
Next, we use algorithm of Sec.~\ref{sec:algorithm} to obtain
$\psi_\Delta$ throughout $\Omega$, and then multiply by $z^{3/2}$ to
get $h_\Delta$. We define the `distance' between two arbitrary
functions on the brane as
\begin{equation}\label{eq:inner product}
    \dlangle f_1 - f_2 \drangle_b =
    \left[ \frac{1}{\eta_f - \eta_i} \int\limits_{\di\Omega_b}
    (f_1-f_2)^2 \right]^{1/2},
\end{equation}
which can be thought of as the root-mean-square (RMS) deviation
between $f_1(\eta)$ and $f_2(\eta)$. Eq.~(\ref{eq:convergence
expectation}) then predicts
\begin{equation}
    \sigma_b(\Delta) \equiv \dlangle h_\Delta - h_\text{exact} \drangle_b
    \sim \text{const.} \times \Delta^2,
\end{equation}
provided that
\begin{equation}
    \Delta \ll r_c \text{ and } k\Delta \ll 1.
\end{equation}
In the second inequality, we have used that $V(z) =
\mathcal{O}(k^2)$ throughout most of $\Omega$.  Note that because
de Sitter branes have inertial trajectories, $r_c = \infty$ and
the first inequality is trivially satisfied.

We have calculated $\sigma_b$ for a wide variety of simulations of
de Sitter brane scenarios and show our results in Fig.~\ref{fig:dS
converge}.  Each curve represents families of simulations where
$\Delta$ is varied, but all other parameters are kept fixed.  We
see that $\sigma_b$ is indeed proportional to $\Delta^2$ for
$\Delta$ sufficiently small, which allows us to draw two
conclusions: First, the numeric solution is indeed approaching the
exact solution in the limit of $\Delta \rightarrow 0$; and second,
the rate of convergence is quadratic.  Notice that this quadratic
convergence sets in for lower values of $\Delta/z_*$ as
$\epsilon_*$ is increased; this is because of
Eq.~(\ref{eq:k-epsilon connection}), which says that $kz_*$ scales
like $\epsilon_*$ when $\epsilon_*$ is large.  Hence, in order to
satisfy $k\Delta \ll 1$, $\Delta$ must approach 0 as $\epsilon_*
\rightarrow \infty$.
\begin{figure}
\begin{center}
\includegraphics{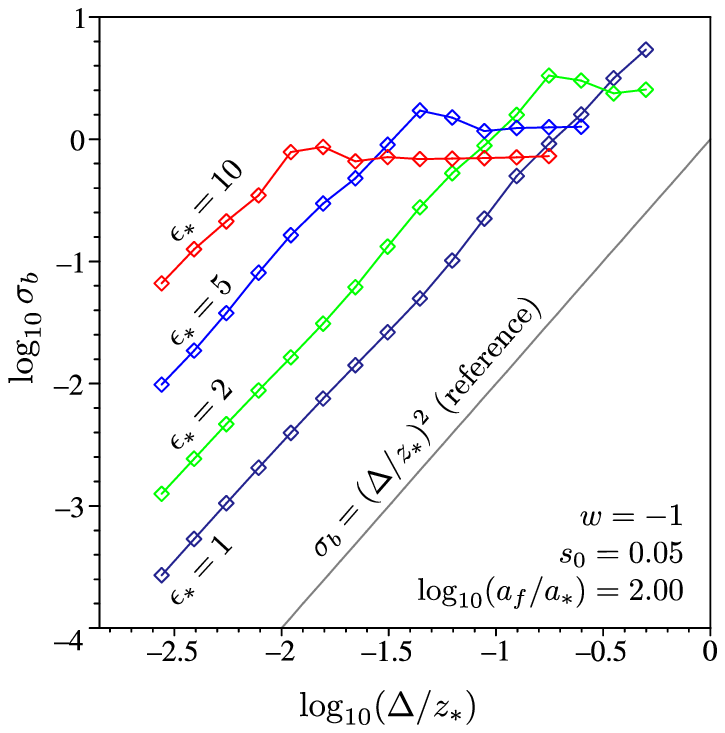}\\
\caption{(Colour online.) RMS deviation between exact and numeric
solutions $\log_{10}\sigma_b$ versus $\log_{10}(\Delta/z_*)$ for
several simulations of tensor fluctuations around inertial (de
Sitter) branes. Quadratic convergence, $\sigma_b(\Delta) \propto
\Delta^2$, is evidenced for $\Delta$ sufficiently small, indicating
our numeric algorithm is both stable and accurate under these
conditions.}\label{fig:dS converge}
\end{center}
\end{figure}

\subsection{Non-inertial branes}

The code tests in the last section were for manifestly
non-accelerating branes, so one might reasonably worry about the
reliability of our algorithm for more complicated brane
trajectories.  However, one cannot test the numerics in the same
manner as before, precisely because no convenient exact solution
to the wave equation is known for an accelerating brane.  Indeed,
this was one of the main motivations of developing the algorithms
of Sec.~\ref{sec:numeric}.

\begin{figure}
\begin{center}
\includegraphics{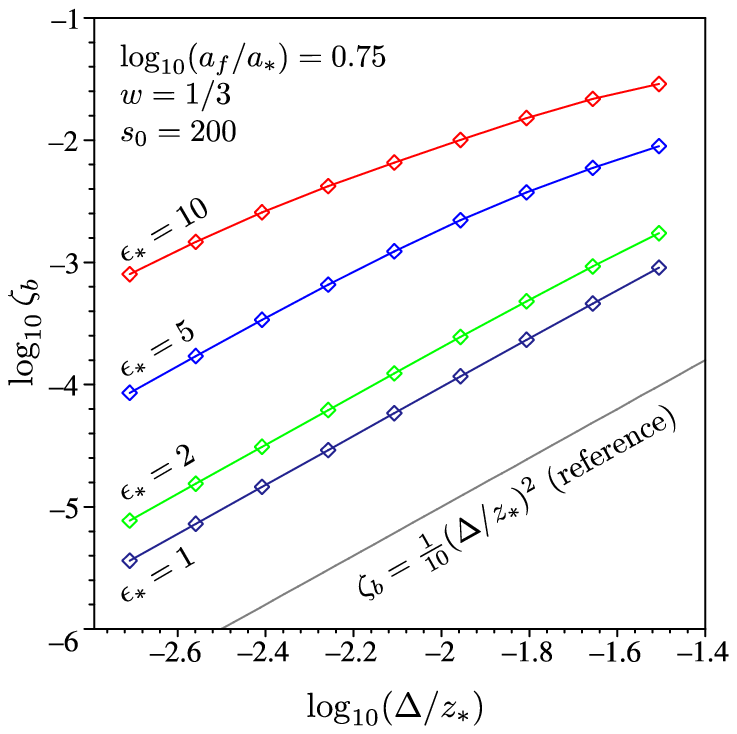}\\
\caption{(Colour online.) The convergence statistic
$\log_{10}\zeta_b$ versus $\log_{10}(\Delta/z_*)$ for several
simulations of GWs about noninertial (radiation-dominated) branes.
As in Fig.~\ref{fig:dS converge}, quadratic convergence,
$\zeta_b(\Delta) \propto \Delta^2$, is apparent as $\Delta
\rightarrow 0$.}\label{fig:noninertial converge}
\end{center}
\end{figure}

In the absence of an exact solution, we can test for the convergence
of our numeric results, but not the accuracy.  In other words, we
can confirm that our results stably approach some limit (in the
Cauchy sense) as $\Delta \rightarrow 0$, but we do not know if it is
the right answer.  A convergence test can be formulated as follows:
Keeping everything else constant, we run our code once with accuracy
$\sqrt{2}\Delta$ and again with accuracy $\Delta$ (the second run
will take around twice as much CPU time as the first). Then, we can
define the RMS discrepancy on the brane between the two runs as
\begin{equation}
    \zeta_b(\Delta) =
    \dlangle h_\Delta - h_{\Delta/\surd 2} \drangle_b.
\end{equation}
As in the last section, eq.~(\ref{eq:convergence expectation})
predicts that this statistic should obey
\begin{equation}
    \zeta_b(\Delta) = \text{const.} \times \Delta^2, \quad \Delta \ll
    r_c \text{ and } k\Delta \ll 1.
\end{equation}
In Fig.~\ref{fig:noninertial converge}, we plot $\zeta_b$ versus
$\Delta$ for a number of different non-inertial brane trajectories
corresponding to radiation dominated brane universes.  The initial
condition for these simulations is simply
\begin{equation}
    h(\di\Omega^-) = 1.
\end{equation}
Note that, to a large degree, the convergence properties of the
algorithm will be independent of the initial data.  We again see
quadratic convergence in this figure for $\Delta$ small enough, and
that smaller values of $\epsilon_*$ lead to faster convergence.

From this, we can infer that our algorithm is convergent when the
brane is accelerating.  However, our ignorance of the exact solution
means that its accuracy is still (technically) in question.  The
only way to get a handle on the latter is to compare our results
with those independently obtained by some other group/method.  This
is done in Appendix \ref{sec:HKT vs IN}, where we explicitly compare
our results for a particular brane model to those of
\citetalias{Hiramatsu:2004aa}.

\section{Setting the initial condition for bulk fluctuations from
inflation}\label{sec:initial conditions}

Having convinced ourselves that the numeric algorithm developed in
Sec.~\ref{sec:numeric} is returning reliable results, we turn
attention to the physical problem we want to study.  This is the the
evolution of tensor perturbations in the high-energy radiation
regime after the end of inflation in braneworld cosmology. In this
section, we discuss the difficulties associated with determining the
initial conditions of our classical problem from a quantum
mechanical analysis of inflation.

\begin{figure}
\begin{center}
\includegraphics[width=0.95\columnwidth]{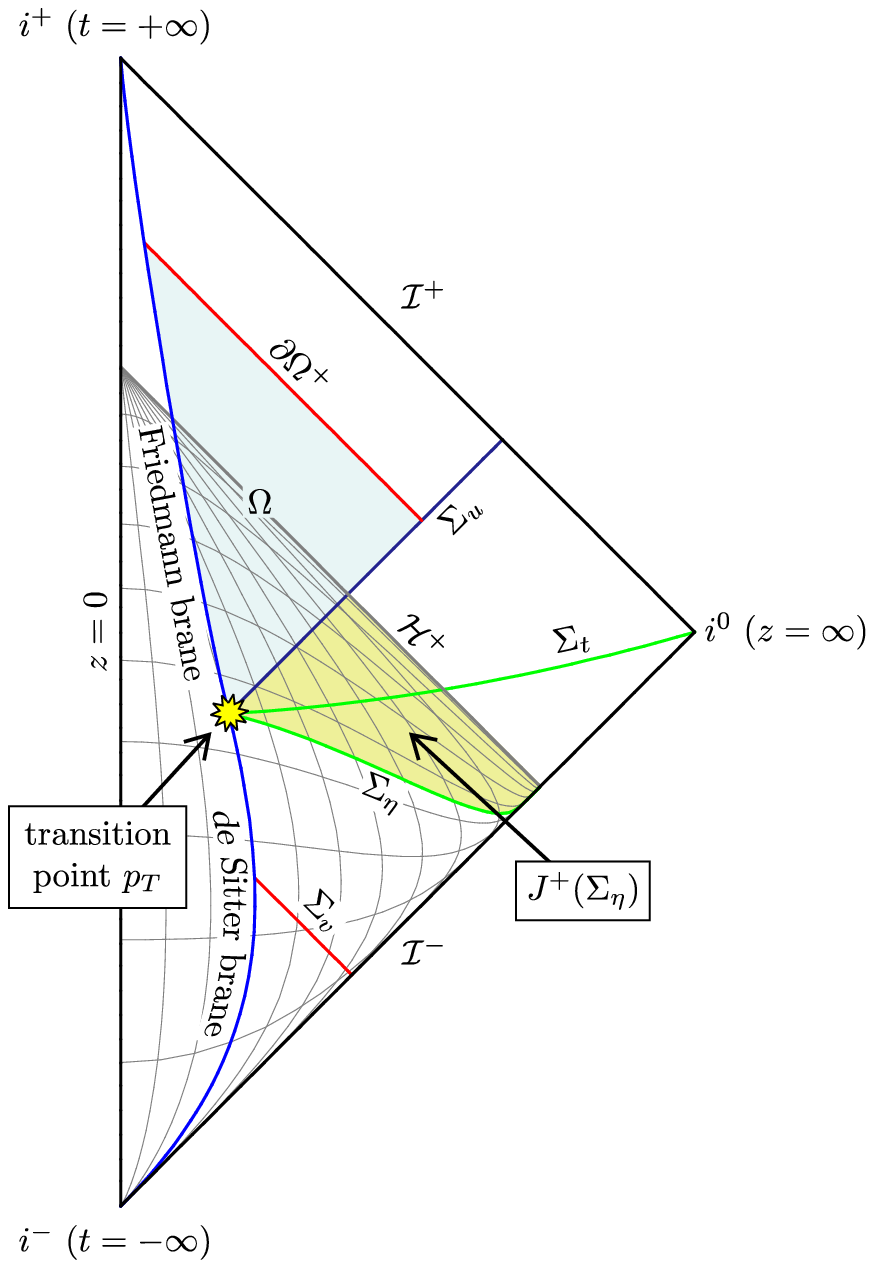}\\
\caption{(Colour online.) Conformal diagram illustrating the causal
structure of a braneworld model of the early
universe}\label{fig:conformal}
\end{center}
\end{figure}

The simplest picture of RS brane cosmology in the early universe
assumes that the brane has some initial phase of pure-de Sitter
inflation followed by a period of radiation-dominated expansion.
These two distinct brane trajectories are smoothly joined at some
transition point $p_T$ in the $(t,z)$ plane.  The situation is
illustrated in Fig.~\ref{fig:conformal} via a conformal diagram.
To generate this plot, we have applied the standard
compactification to the the $(t,z)$ coordinates:
\begin{equation}
    T = \frac{\tanh v + \tanh u}{2}, \quad
    Z = \frac{\tanh v - \tanh u}{2}.
\end{equation}
We see that the brane begins in the vicinity of past timelike
infinity $i^-$, reaches the transition point at some finite $(t,z)$,
and then continues expanding as a Friedmann brane on to future
timelike infinity $i^+$.

Now, the standard paradigm concerning braneworld GW perturbations
is they are generated quantum mechanically during the de Sitter
phase of the expansion. The calculation relies heavily on the
$(\eta,\xi)$ Rindler-like coordinates introduced in
Sec.~\ref{sec:de Sitter code test}, since the exact solution of
the classical equation of motion is known exactly in that patch.
The coordinate lines of this patch have been drawn on
Fig.~\ref{fig:conformal}.  One assumes that $h(\eta,\xi)$ is
described by a quantum field in its vacuum state, as measured by
observers travelling on $\xi =$ constant slices. We denote this
`de Sitter invariant vacuum' by $|0\rangle_\eta$.  Then, one
simply evaluates the vacuum expectation values of the squared
amplitude of the various mode functions to see how they are
magnified during inflation; for example, the zero mode amplitude
is
\begin{equation}
    {}_\eta\langle 0 | \hat\phi^2_0 | 0 \rangle_\eta =
    |\phi_0(\eta,\xi)|^2 = \text{function of $\eta$ only}.
\end{equation}
That is, the evolution of the quantum fluctuations is determined
by the amplitude of the classical mode functions, which are known
exactly.  In this way, \citet{Langlois:2000ns} have shown that
while the amplitude of the zero mode $\phi_0$ grows during
inflation, the amplitudes of the so-called `massive modes'
$\phi_\nu$ are suppressed, which leads to a scale-invariant
spectrum that agrees with the 4-dimensional result,
\begin{eqnarray}\nonumber
    {}_\eta\langle 0 | \hat h^2 | 0 \rangle_\eta & = &
    |\phi_0(\eta)|^2 \\ & \xrightarrow[\,0\,]{\eta} & \frac{ C^2(\sinh\xi_b)\sinh^2\xi_b}{2 k^3\ell^3} \equiv \mathcal{C}_i^2(k).
\end{eqnarray}
This quantum calculation suggests that the appropriate
post-inflationary initial condition for classical GW calculations
is
\begin{equation}\label{eq:inflation IC}
    \text{inflation:} \quad h(\Sigma_\eta) = \text{const.},
\end{equation}
where $\Sigma_\eta$ is an $\eta =$ constant hypersurface that
intersects the brane `at the end of inflation'
(\emph{cf.}~Fig.~\ref{fig:conformal}), which for our purposes can
be taken as the transition point $p_T$.

But there is problem with this picture:  In order to have a
well-defined Cauchy problem for the fluctuations $h(t,z)$ from the
transition time to the infinite future, we must have initial data on
the entire $\Sigma_u$ hypersurface, which is the $u =$ constant line
running from $p_T$ to $\mathcal{I}^+$.  Now, by specifying initial
data on $\Sigma_\eta$, we can immediately use the bulk wave equation
to obtain the field value in it's causal future
$J^+(\Sigma_\eta)$.\footnote{This is possible to do analytically,
since the general solution of the bulk wave equation is known in
closed form, and evolution in $J^+(\Sigma_\eta)$ proceeds
independently of the brane boundary condition.  This holds for any
spacelike hypersurface whose causal future does not intersect the
brane.}  But this is not sufficient to determine the value of the
field on $\Sigma_u$; it is apparent from Fig.~\ref{fig:conformal}
that
\begin{equation}
    \Sigma_u \nsubseteq \bar{J}^+(\Sigma_\eta).
\end{equation}
Here, an overbar indicates the closure of a set.  The reason that
$\Sigma_u$ does not lie in the future domain of dependence of
$\Sigma_\eta$ is the existence of a Cauchy horizon $\mathcal{H}^+$
in the $(\eta,\xi)$ coordinates. In Figs.~\ref{fig:dS geometry} and
\ref{fig:conformal}, this is the future boundary of the portion of
the $(t,z)$ plane covered by the de Sitter coordinates.  Note that
this problem is not in general mitigated by choosing to model the
brane in a finite computational domain $\Omega$; as in
Fig.~\ref{fig:conformal}, one can draw many examples where
$\di\Omega^- \nsubseteq \bar{J}^+(\Sigma_\eta)$.

There are several ways to get around the fact that
$\bar{J}^+(\Sigma_\eta)$ is too small. In the literature, several
authors note that the wave equation and boundary condition
(\ref{eq:RS eqns}) can be solved approximately for extremely long
wavelengths:
\begin{equation}
    k \rightarrow 0, \quad h(t,z) \sim \text{constant for all
    $(t,z)$}.
\end{equation}
The conclusion that one draws is that, in the limit, the inflation
initial condition (\ref{eq:inflation IC}) is `consistent' with
setting $h$ = constant on some other, more suitable hypersurface.
Consequently, \citetalias{Hiramatsu:2004aa} have adopted the initial
conditions
\begin{equation}\label{eq:HKT IC}
    \text{\citetalias{Hiramatsu:2004aa}:} \quad h(\Sigma_t) =
    \text{const.}, \quad h_{,t}(\Sigma_t) = 0,
\end{equation}
where $\Sigma_t$ is the hypersurface running from the transition
point $p_T$ to spatial infinity, as shown in
Fig.~\ref{fig:conformal}. This is sufficient to specify the Cauchy
evolution of $h$, since
\begin{equation}
    \Sigma_u \subset \bar{J}^+(\Sigma_t).
\end{equation}
Note that if the brane were static, this initial data would
reproduce the zero mode of the original RS model:
\begin{equation}
    \varphi_0(t,z) = \text{const.} \times e^{-ikt}.
\end{equation}
Of course, this is not a unique prescription;
\citetalias{Ichiki:2004sx} have instead elected to enforce:
\begin{equation}\label{eq:IN IC}
    \text{\citetalias{Ichiki:2004sx}: } \quad h(\Sigma_u) = \text{const.};
\end{equation}
i.e., they have set the perturbation equal to a constant on the
initial null hypersurface.  Both groups acknowledge that these
initial condition are somewhat \emph{ad hoc}; when $k = 0$ they are
only consistent with the inflation initial condition
(\ref{eq:inflation IC}) if additional data is specified on
$\mathcal{I}^-$, and they are not even consistent with one another
for $k > 0$.  We will explicitly contrast the
\citetalias{Hiramatsu:2004aa} and \citetalias{Ichiki:2004sx} initial
conditions in Sec.~\ref{sec:HKT vs IN}.

A separate approach comes from treating the quantum inflationary
calculation differently.  \citet{Gorbunov:2001ge} consider a
`junction model' that has de Sitter and Minkowski branes attached to
one another at a non-smooth transition point.  They assume that the
GWs are in the de Sitter invariant vacuum $|0\rangle_\eta$ in the
infinite past, which implies quantum particle creation when the
brane's trajectory changes abruptly. At the end of the day, then
derive the spectrum of GWs as seen by RS observers travelling
orthogonal to $t =$ constant slices. At first glance, this would
seem to be ideal since the final amplitudes are given on $\Sigma_t$;
furthermore, the actual derived spectrum is dominated by the zero
mode, and is hence consistent with the \citetalias{Hiramatsu:2004aa}
initial condition (\ref{eq:HKT IC}).  However, some caution is
required here, since the calculation essentially involves
decomposing the quantum states preferred by $z =$ constant observers
in terms of those preferred by $\xi =$ constant observers.  But the
latter basis is only really defined to the past of the Cauchy
horizon, so we may legitimately worry if the field amplitude on the
entirety of $\Sigma_t$ is fixed in this approach.  (This issue is
clearly beyond the scope of the current work.)

Recently, \citet{Kobayashi:2005dd} have modified the
\citeauthor{Gorbunov:2001ge}~calculation by smoothly joining the
initial and final phases by a radiation-dominated brane.  In order
to calculate the Bogliobov coefficients governing particle
creation, they follow the usual practice of fixing the field value
in the future (on an $v$ = constant slice) and calculation what it
is in the past (on an initial $v$ = constant slice like $\Sigma_v$
in Fig.~\ref{fig:conformal}).  They found a subdominant amount of
energy $\lesssim 10 \%$ is stored in the Kaluza-Klein modes at the
end of the quantum regime.  \citet{Kobayashi:2006pe} subsequently
repeated the calculations with an improved numerical scheme, which
resulted in more robust predictions.

Finally, we note one additional strategy for specifying initial
data; namely, to enforce boundary conditions on $\scriminus$.
Imposition such an additional constraint is enough to fix the
problems associated with the inflation initial condition
(\ref{eq:inflation IC}), since
\begin{equation}
    \Sigma_u \subset \bar{J}^+(\Sigma_\eta \cup \mathcal{I}^{-}).
\end{equation}
The real question is: what is the appropriate condition to choose?
A physically sensible minimal condition would be that there is no
radiative flux entering our patch of AdS space through
$\scriminus$.\footnote{One could argue that, in some sense, the
potential enforces this condition for us:  Far from the brane,
$\psi$ behaves like a field of mass $k$ propagating in free space.
Hence, wave packets of $\psi$ must originate at $i^-$, and not
from past null infinity.  In other words, wave packets on
$\mathcal{I}^-$ should never reach the brane, and therefore it
seems that initial data on $\mathcal{I}^-$ can have no relevance
to the observed GW spectrum. However, this conclusion relies both
on the properties of the potential (and hence fails for $k=0$),
and an eikonal approximation. Here, we are interested in making
statements independent of the potential and any approximations,
which means that specification of initial data on $\mathcal{I}^-$
is logically distinct from the specification of data on
$\Sigma_\eta$, $\Sigma_v$, etc.  Also, from a practical point of
view, it is of little use to assume that our solution is
independent of the field configuration on $\mathcal{I}^-$; our
numeric code crashes unless data is specified on an initial
$u$-slice, no matter how far in the past it is.} It would be
interesting to examine this type of prescription in detail, but
such a project is beyond the scope of the current work.

\section{HKT versus IN formulations}\label{sec:HKT vs IN}

In the previous section, we saw several different types of initial
conditions that one could use for numerical simulations of bulk
GWs in the post-inflationary epoch of brane cosmology.  In this
section, we explicitly compare the GWs generated by the
\citetalias{Hiramatsu:2004aa} and \citetalias{Ichiki:2004sx}
initial conditions.  It is useful to first identify the key
qualitative features of the radiation, and to do this we
concentrate on the \citetalias{Hiramatsu:2004aa} case. For
simplicity, we take the \citetalias{Hiramatsu:2004aa} condition to
be $h = 1$ and $\di_t h = 0$ on an initial $t$ slice in actual
calculations. Labeling the initial time as $t = 0$, we can write
down the \emph{exact} solution of the wave equation (\ref{eq:RS
eqns}) inside the causal future of $\Sigma_t$:
\begin{equation}
    h(t,z) = \cos(kt) \text{ for all } (t,z) \in
    \bar{J}^+(\Sigma_t).
\end{equation}
In particular, this can be evaluated on $\Sigma_u$ to give us
initial data on $\di\Omega^-$,
\begin{equation}\label{eq:RS zero null projection}
    h(\di\Omega^-) = \cos \tfrac{1}{2} k(v + u_i),
\end{equation}
where $u_i$ is the $u$-coordinate of the intersection of the brane
with the initial time slice.

\begin{figure}
\begin{center}
\subfigure[\,\,Bulk profile]%
{\includegraphics{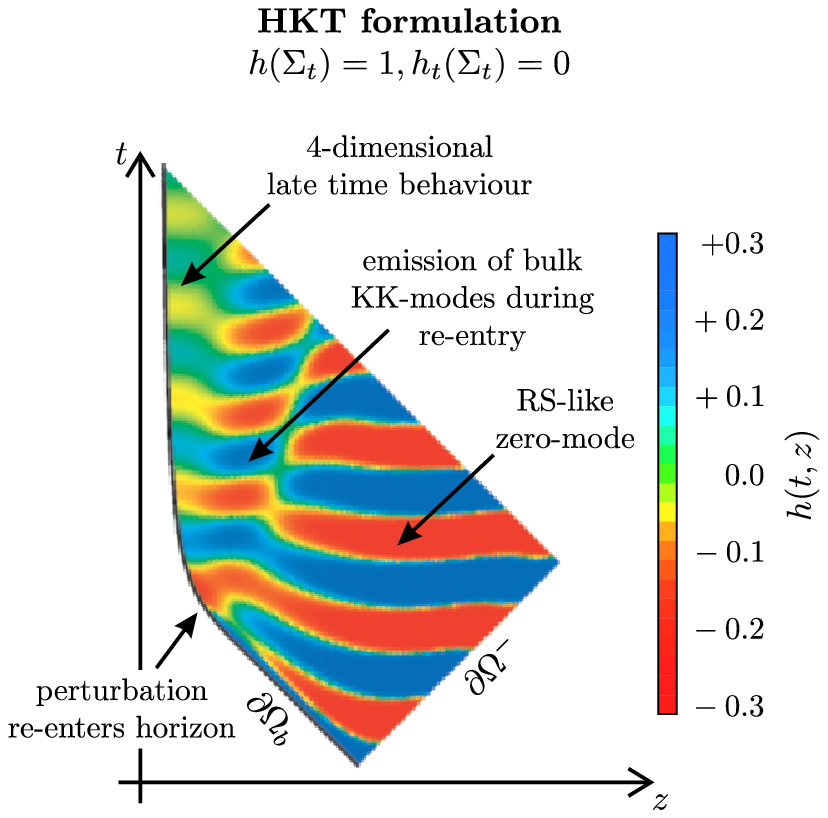}\label{fig:HKT bulk}}\\
\subfigure[\,\,Brane profile]%
{\includegraphics{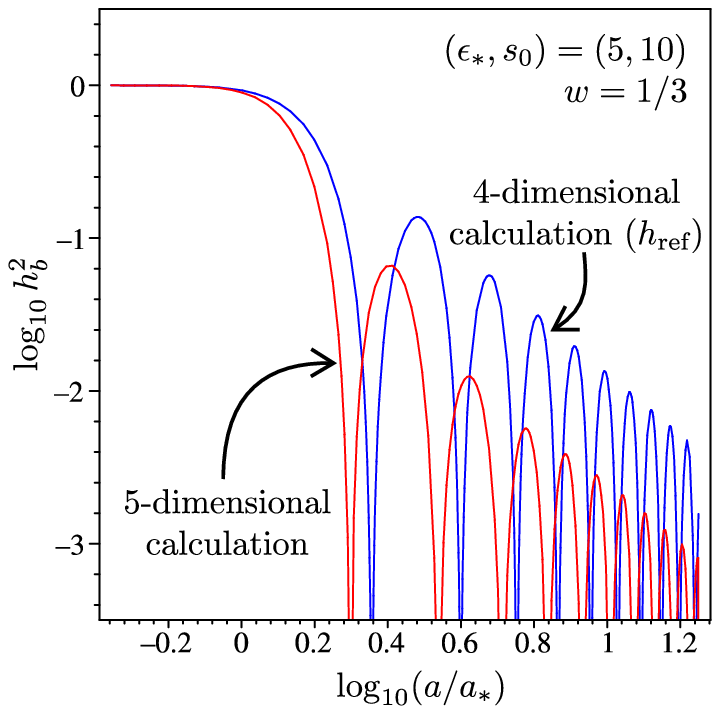}\label{fig:HKT brane}}%
\caption{(Colour online.) Results of a typical integration using
the \citetalias{Hiramatsu:2004aa} initial condition ($\Delta/z_* =
2^{-10} \sim 10^{-3}$). In Fig.~\ref{fig:HKT brane}, we have drawn
what the brane GW signal would be if the bulk were neglected;
i.e., if one solved the 4-dimensional master equation (\ref{eq:4D
master equation 1}) with a modified expansion rate given by
(\ref{eq:ordinary Friedmann}).  Note that we have enforced the
initial condition that $h_b = h_\text{ref}$ for $a \ll a_*$ on the
reference wave.}\label{fig:HKT example}
\end{center}
\end{figure}
In Fig.~\ref{fig:HKT example}, we plot the results of our numeric
calculation for a particular case involving a radiation dominated
brane. The key features are as follows: In the bulk, the waveform
appears to retain the character of the Randall-Sundrum zero mode
far away from the brane; i.e., it is constant on constant $t$
slices. However, this symmetry is broken closer to the brane by
the motion of the boundary, resulting in rich and intriguing
dynamics.  On the brane, the GW amplitude remains constant for $a
\lesssim a_*$, and then begins to oscillate and decay.
Asymptotically, one can easily confirm that the numeric waveform
goes like
\begin{equation}\label{eq:5D asymptotic}
    h_b \xrightarrow[\,\infty\,]{a} \frac{\mathcal{C}_b(k)}{a}
    \cos \left( \omega_0 \frac{a}{a_*} + \varsigma \right) = \frac{\mathcal{C}_b(k)}{a} \cos(k
    \eta + \varsigma),
\end{equation}
where $\mathcal{C}_b(k) > 0$ is the \emph{expansion-normalized
characteristic amplitude}, $\varsigma$ is a phase, and
\begin{equation}
    \omega_0 = \sqrt{1+\tfrac{1}{2}\epsilon_*}.
\end{equation}
This asymptotic form can be understood by considering a
fictional 4-dimensional universe that undergoes the same expansion
(\emph{cf.}~Eq.~\ref{eq:Friedmann}) as our `real' brane universe.
Tensor fluctuations in such a model obey
\begin{equation}\label{eq:4D master equation 1}
    \left( \frac{d^2}{d\eta^2} + 2Ha \frac{d}{d\eta} + k^2 \right)
    h_\text{ref} = 0,
\end{equation}
where the `ref' subscript indicates the amplitude in our reference
4-dimensional universe.  At late times, $H\ell \sim
\sqrt{2\epsilon_*} (a_*/a)^{2}$ for a radiation dominated brane,
leading to
\begin{equation}\label{eq:4D asymptotic}
    h_\text{ref}(a) \xrightarrow[\,\infty\,]{a} \frac{\mathcal{C}_\text{ref}(k)}{a}
    \cos(k\eta + \varsigma_\text{ref}), \quad \mathcal{C}_\text{ref}(k) > 0.
\end{equation}
Hence, the late time behaviour of $h_b$ from the numeric
simulations matches that of the 4-dimensional effective
calculation.  However, the amplitude of $h_b$ relative to
$h_\text{ref}$ is somewhat diminished, as seen Fig.~\ref{fig:HKT
brane}.  Physically, this has to do with the brane motion inducing
the radiative loss of GW energy into the bulk, as depicted by the
bulk gravity wave profile seen in Fig.~\ref{fig:HKT bulk}.

\begin{figure}
\begin{center}
\includegraphics{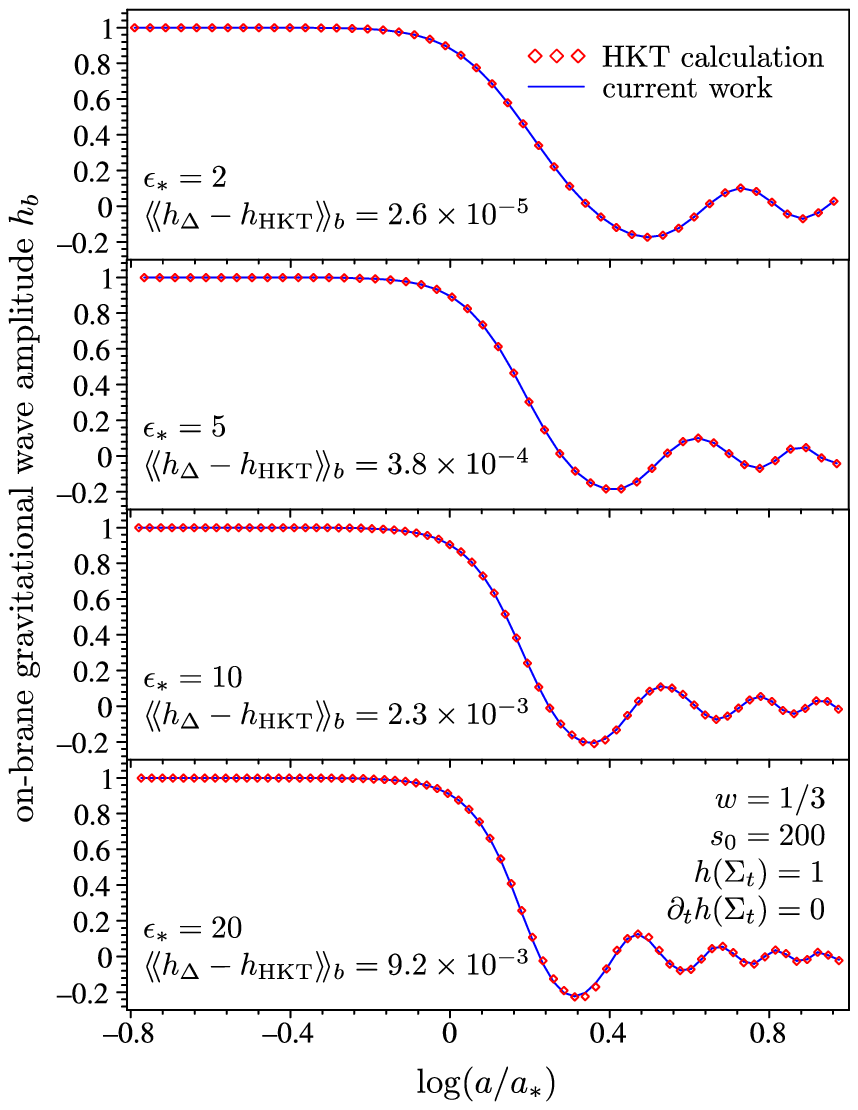}\\
\caption{(Colour online.) Direct comparison of numerical results
obtained by \citetalias{Hiramatsu:2004aa} and the current work
($\Delta/z_* = 2^{-10} \sim 10^{-3}$). In each panel, the RMS
deviation between the two profiles is given.}\label{fig:Hiramatsu
compare}
\end{center}
\end{figure}
In Fig.~\ref{fig:Hiramatsu compare}, we plot our $h_b$ predictions
for several additional cases using the \citetalias{Hiramatsu:2004aa}
initial condition. For comparison, we also plot the actual numeric
results obtained by \citetalias{Hiramatsu:2004aa} for the same
cases. The agreement between the two independent calculations is
visually excellent, with both sets of data lining up very well, but
not perfectly. We can quantify the level of agreement by calculating
the RMS discrepancies $\dlangle h_\Delta - h_\text{HKT} \drangle_b$,
which are written directly on the figures.  We see that $\dlangle
h_\Delta - h_\text{HKT} \drangle_b \lesssim 10^{-2}$ for the cases
shown, which can be interpreted as the average absolute deviation
between the two results.  This is acceptably small, and we can
conclude that the two simulations agree to within reasonable
tolerances.

\begin{figure}
\begin{center}
\subfigure[\,\,Bulk
profile]{\includegraphics[width=\columnwidth]{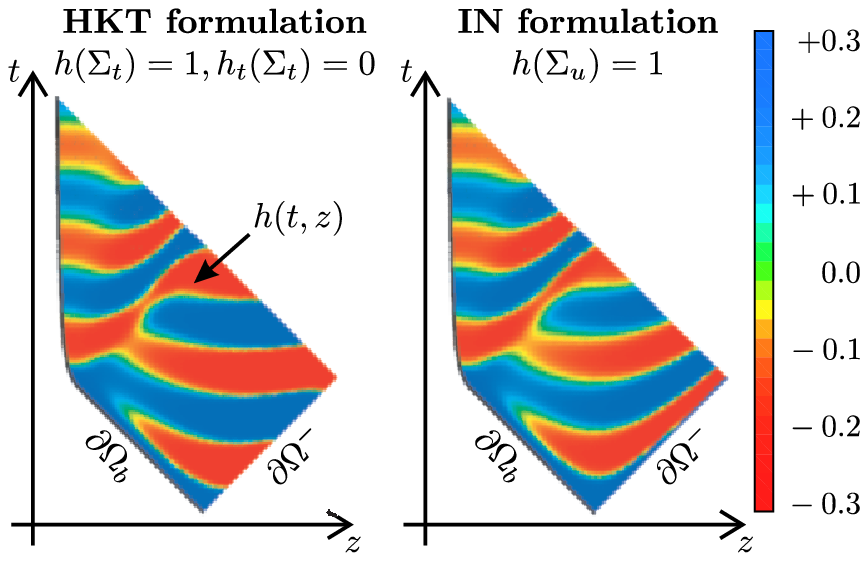}\label{fig:stiff
matter bulk}}
\\ \subfigure[\,\,Brane
profile]{\includegraphics{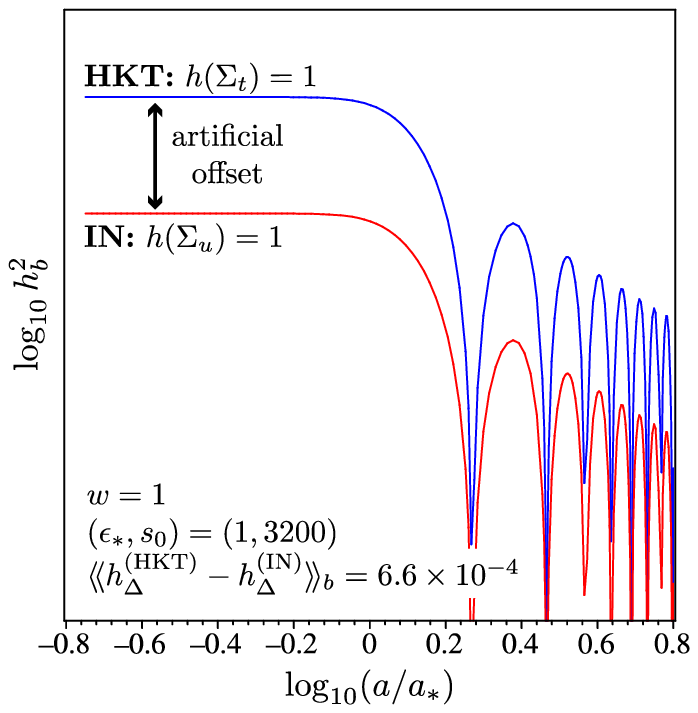}\label{fig:stiff
matter brane}} \caption{(Colour online.) GW amplitudes
$h_\Delta(t,z)$ for a `stiff matter' brane ($w = 1$) using the
\citetalias{Hiramatsu:2004aa} and \citetalias{Ichiki:2004sx}
initial conditions.}\label{fig:HKT vs IN 1}
\end{center}
\end{figure}

Now, we turn our attention to the \citetalias{Ichiki:2004sx}
formulation. In Fig.~\ref{fig:HKT vs IN 1}, we show the radiation
patterns around a $w = 1$ `stiff matter' brane generated by the
\citetalias{Hiramatsu:2004aa} and \citetalias{Ichiki:2004sx}
initial conditions, respectively.  The two bulk waveforms in
Fig.~\ref{fig:stiff matter bulk} are similar to one another, but
exhibit some clear differences, especially near the initial data
surface $\di\Omega^-$.  But the the brane profiles shown in
Fig.~\ref{fig:stiff matter brane} are virtually identical to one
another.  Indeed, the RMS discrepancy between the two results is
$\sim 10^{-4}$, which is quite small.

Is it possible that the remarkable on-brane agreement between the
\citetalias{Hiramatsu:2004aa} and \citetalias{Ichiki:2004sx}
formulations is due to a serendipitous choice of parameters?  In
Fig.~\ref{fig:HKT vs IN 2}, we plot the discrepancy (as defined by
the inner product Eq.~\ref{eq:inner product}) between the brane
amplitude generated by the two different initial conditions as a
function of the initial wavelength of the perturbation $s_0$.  In
this plot, the brane is radiation-dominated.  We see the limiting
behaviour
\begin{equation}
    \dlangle h_\Delta^\text{(HKT)} - h_\Delta^\text{(IN)} \drangle_b
    \xrightarrow[\,\infty\,]{s_0} 0;
\end{equation}
i.e., as the initial data surface $\di\Omega^-$ is moved further and
further into the past (cf.~Fig.~\ref{fig:domain}), the level of
agreement between the two boundary conditions increases.  This is
reminiscent of a result obtained by \citetalias{Hiramatsu:2004aa}:
Their on-brane waveforms showed very little dependence on $s_0$,
provided that $s_0$ was large enough.
\begin{figure}
\begin{center}
\includegraphics{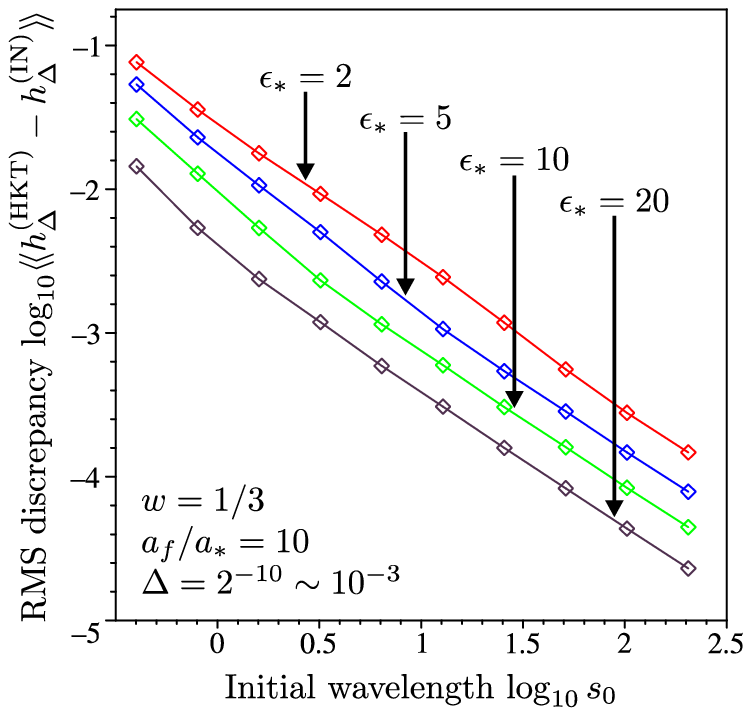}
\caption{(Colour online.) RMS discrepancy in the $h_b$ waveforms
generated by the \citetalias{Hiramatsu:2004aa} and
\citetalias{Ichiki:2004sx} initial conditions versus the initial
wavelength $\log_{10}s_0$ of the perturbation.}\label{fig:HKT vs
IN 2}
\end{center}
\end{figure}

\begin{figure}
\begin{center}
\includegraphics{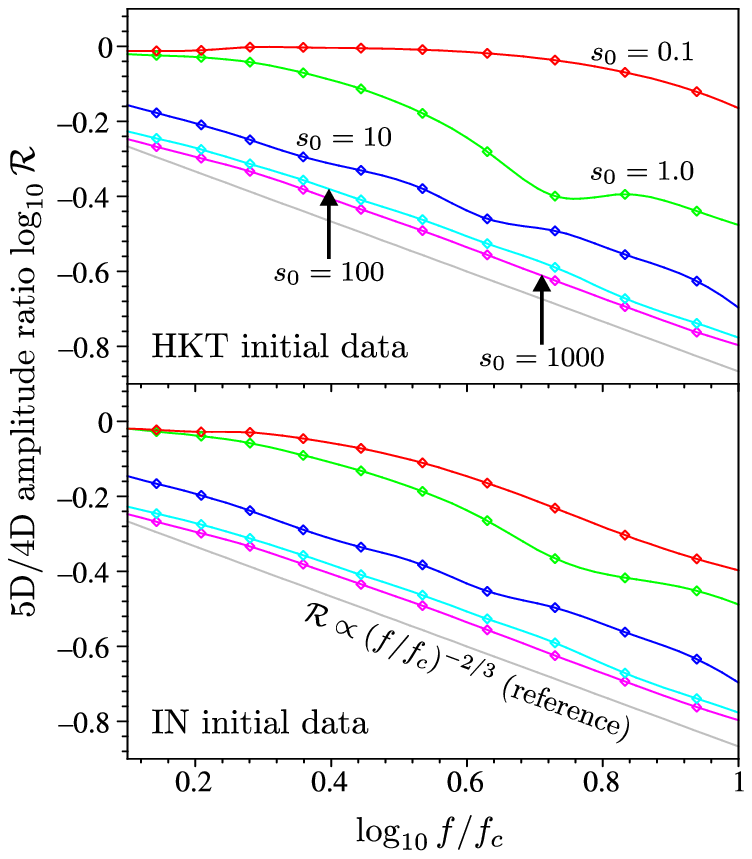}
\caption{(Colour online.) The ratio $\mathcal{R}$ of the late-time
amplitude of GWs on the brane obtained from our simulations to
that of the reference wave $h_\text{ref}$. In all cases, we have
assumed radiation domination $w = 1/3$ and evaluated the ratio
when $a = 20a_*$, or equivalently $\rho/\lambda \lesssim 2 \times
10^{-4}$.}\label{fig:amplitude ratio}
\end{center}
\end{figure}

As described in Appendix \ref{sec:GW background}, we need to know
\begin{equation}
    \mathcal{R} = \frac{\mathcal{C}_b}{\mathcal{C}_\text{ref}} =
    \frac{\text{late time amplitude of $h_b$}}{\text{late time amplitude
    of $h_\text{ref}$}}
\end{equation}
in order to predict the observable spectral density of GWs today.
Here, we will take the post-inflationary epoch to be purely
radiation-dominated.  By `late time', we mean that the ratio
should be evaluated in the low energy regime of the cosmological
evolution:
\begin{equation}
    1 \gg \frac{\rho}{\lambda} = \epsilon_* \left(\frac{a_*}{a}
    \right)^4.
\end{equation}
For practical calculations, we measure the amplitude ratio when $a
= 20a_*$ and limit $\epsilon_* \leq 32$.  In
Fig.~\ref{fig:amplitude ratio} we plot $\mathcal{R}$ as a function
of $f/f_c$ for the \citetalias{Hiramatsu:2004aa} and
\citetalias{Ichiki:2004sx} initial conditions and several choices
of $s_0$; where $f$ is the present-day frequency of the simulated
wave, and $f_c$ is the present-day frequency of a mode that
re-entered the horizon when $H\ell = 1$ (cf.~Eq.~\ref{eq:f/f_c
ratio}).  Qualitatively, we see that for $s_0 \lesssim 1$, there
is a discernable difference in the $\mathcal{R}$ predictions from
the two initial conditions.  However for $s_0 \gg 1$, the ratios
are identical to one another, and satisfy
\begin{equation}
    \mathcal{R} \propto (f/f_c)^{-2/3}, \quad f \gg f_c.
\end{equation}
When this result is combined with (\ref{eq:delta_T from
inflation}) and (\ref{eq:spectral density}), we find that the
present day spectral density of gravitational radiation obeys
\begin{equation}
    \Omega_\text{GW} \propto \left( \frac{f}{f_c} \right)^0, \quad
    f \gg f_c;
\end{equation}
i.e., we get a flat GW spectrum in the high-frequency limit.  This
essentially re-confirms the main result of
\citetalias{Hiramatsu:2004aa}, but with the twist that it also
holds for the initial data prescription favoured by
\citetalias{Ichiki:2004sx}.  However, the latter group claims the
spectrum is red at high frequencies: $\Omega_\text{GW} \propto
(f/f_c)^{-0.46}$.  The source of tension between this and the
current result is unclear.

\section{Generic initial data}\label{sec:generic initial data}

In the previous subsection we saw that if our initial data surface
was set far enough into the past, the on-brane waveforms were
insensitive to the choice between \citetalias{Hiramatsu:2004aa}
and \citetalias{Ichiki:2004sx} initial conditions.  However, it is
clear that either alternative represents a fairly restrictive
choice of initial data. In this subsection, we will explore the
extent to which $h_b$ is indifferent to more arbitrary choices of
$h(\di\Omega^-)$.  Throughout this entire section, we specialize
to radiation dominated $w = 1/3$ models.

\subsection{Basis functions}

Generically, any initial data on $\di\Omega^-$ may be decomposed
as
\begin{equation}
    h(\di\Omega^-) = h_i(v) = \frac{1}{2\pi} \int\limits_{-\infty}^\infty
    d\mu \, \mathcal{A}(\mu) e^{i\mu k(v+u_i)/2},
\end{equation}
where the Fourier amplitude $\mathcal{A}(\mu)$ is arbitrary.  For
example, the \citetalias{Hiramatsu:2004aa} condition is recovered if
\begin{equation}\label{eq:HKT spectrum}
    \mathcal{A}(\mu) = \pi \left[ \delta(\mu-1) + \delta(\mu+1)
    \right],
\end{equation}
while the \citetalias{Ichiki:2004sx} condition follows from
\begin{equation}
    \mathcal{A}(\mu) = 2\pi \delta(\mu).
\end{equation}
Let us now define $\chi_\mu(\Omega)$ to be a solution of the wave
equation such that $\chi_\mu(\di\Omega^-) = e^{i\mu k(v+u_i)/2}$.
Then,
\begin{equation}\label{eq:h decomposition}
    h(\Omega) = \frac{1}{2\pi} \int\limits_{-\infty}^\infty
    d\mu \, \mathcal{A}(\mu) \chi_\mu(\Omega);
\end{equation}
i.e., if we have knowledge of the `basis functions' $\chi_\mu$, we
can write down the solution for $h$ corresponding to arbitrary
initial data.  Our present goal is to numerically calculate
$\chi_\mu$ for various situations and gain some intuition about
its qualitative behaviour.

Before proceeding, it is useful to make a few remarks about
$\{\chi_\mu\}$. First and foremost, this basis is in no way
preferred or special, it is merely convenient.  Its definition is
intimately tied to the choice of the initial null hypersurface;
hence, if $\di\Omega^-$ is moved, we get a different basis.  We do
not expect $\{\chi_\mu\}$ to satisfy any type of orthogonality
relationship over $\Omega$; indeed, we will not even specify what
the appropriate inner product might be.

An important property of $\{\chi_\mu\}$ is the physical
interpretation of the real and imaginary parts of the basis
functions.  If $\bm{p_i}$ is the point where the brane $\di\Omega_b$
and the initial data hypersurface $\di\Omega^-$ intersect, the real
and imaginary parts of $\chi_\mu$ satisfy
\begin{equation}
    \text{Re} \, \chi_\mu(\bm{p_i}) = 1,
    \quad \text{Im} \, \chi_\mu(\bm{p_i}) = 0.
\end{equation}
Hence, the two independent components of any given $\chi_\mu$
represent distinct physical possibilities: Either the initial
brane amplitude is nonzero or not.  From a 4-dimensional point of
view, $\text{Re}\,\chi_\mu$ represents a superhorizon perturbation
whose non-zero amplitude is frozen until re-entry. On the other
hand, $\text{Im} \, \chi_\mu$ is a perturbation that exists
`entirely in the bulk' initially; its appearance on the brane
after the end of inflation would be mysterious to an observer
unaware of the extra dimension.  In the special case of $\mu = 1$,
it is easy to confirm that the imaginary part of $\chi_\mu$
satisfies the following initial conditions on the initial $t =$
constant surface:
\begin{equation}
    \text{Im}\,\chi_{\mu=1}(\Sigma_t) = 0,
    \quad \di_t\text{Im}\,\chi_{\mu=1}(\Sigma_t) \ne 0.
\end{equation}
When this is compared to (\ref{eq:HKT IC}), we see that
$\text{Im}\,\chi_{\mu=1}$ satisfies a sort of `complimentary'
\citetalias{Hiramatsu:2004aa} condition.  On the other hand,
$\text{Re}\,\chi_{\mu=1}$ satisfies the
\citetalias{Hiramatsu:2004aa} condition precisely.

Finally, we mention the physical interpretation of the $\mu$
parameter.  If we neglect the brane boundary condition and work in
the limit of $z \rightarrow \infty$, we find that
\begin{equation}\label{eq:approximate chi}
    \chi_{\mu} \approx e^{ik(u-u_i)/2\mu} e^{i\mu k (v+u_i)/2}
\end{equation}
is a solution of (\ref{eq:RS wave}) that satisfies the appropriate
initial conditions on $\di\Omega^-$; i.e., $u = u_i$.  It is
instructive to calculate the flux associated with this `solution':
\begin{eqnarray}\nonumber
    \DOT{j_\mu}{dx} & = & \frac{1}{2i} \left( \chi_{\mu}^* \bm{D}
    \chi_{\mu} - \chi_{\mu} \bm{D}
    \chi_{\mu}^* \right) \bm{\cdot dx} \\
    & \approx & k (\cosh\beta \, dt + \sinh\beta \, dz), \label{eq:flux}
\end{eqnarray}
where we have defined
\begin{equation}
   \tanh\beta = \frac{\mu^2-1}{\mu^2+1}.
\end{equation}
Hence, asymptotically far from the brane and the origin ($z = 0$),
the $\chi_\mu$ basis functions reduce to plane waves traveling with
Lorentz boost parameter $\beta$ with respect to the static $(t,z)$
coordinates.  Note that the definition of $\beta$ implies that modes
with $|\mu| < 1$ have wavevectors pointing towards $z = 0$, while
the modes with $|\mu| > 1$ are traveling away from $z = 0$.  The
\citetalias{Hiramatsu:2004aa} modes $\mu = \pm 1$ represent the
middle ground: they have no relative motion with respect to the
static frame.  Indeed, when $\mu = \pm 1$ we have $\chi_\mu \approx
e^{\pm ikt}$ in the asymptotic region, which are the two independent
phases of the RS zero-mode.

To summarize, in this subsection we have introduced a set of basis
functions in terms of which any square-integrable $h(\di\Omega^-)$
can be decomposed.  It must be stressed that while this choice is
convenient, it is \emph{arbitrary}.  Obviously, if we employed any
other basis, the interpretation of $\mu$ would be quite different;
for example, we could have selected the RS massive mode functions
(evaluated on $\di\Omega^-$) from the static brane case
\cite{Randall:1999vf} as a basis, which is perhaps more suited to
the fact that the bulk is warped.  Having said this, our choice of
$\chi_\mu$ is extremely straightforward to implement numerically,
and we feel that the $\mu$ parameter has an `easy' physical
interpretation: both as the relative frequency of initial data, and
simply related to the Lorentz boost parameter of a plane wave
irradiating the brane from $z = \infty$.

\subsection{Numerical results}\label{sec:numerical results}

In this subsection, we present our numerical results concerning
the evolution of the $\chi_\mu$ basis functions introduced above.
Note that by introducing this basis, we have increased the
parameter space from $(\epsilon_*,s_0,a_f/a_*)$ to
$(\epsilon_*,s_0,a_f/a_*,\mu)$.  Each of these has a continuous
spectrum, so it is very impractical to sample this parameter space
densely. Instead, we will attempt to identify the principal trends
in the waveform behaviour from the simulations, and go on to
rationalize them with some approximate analytical work in the next
subsection.

\begin{figure}
\includegraphics{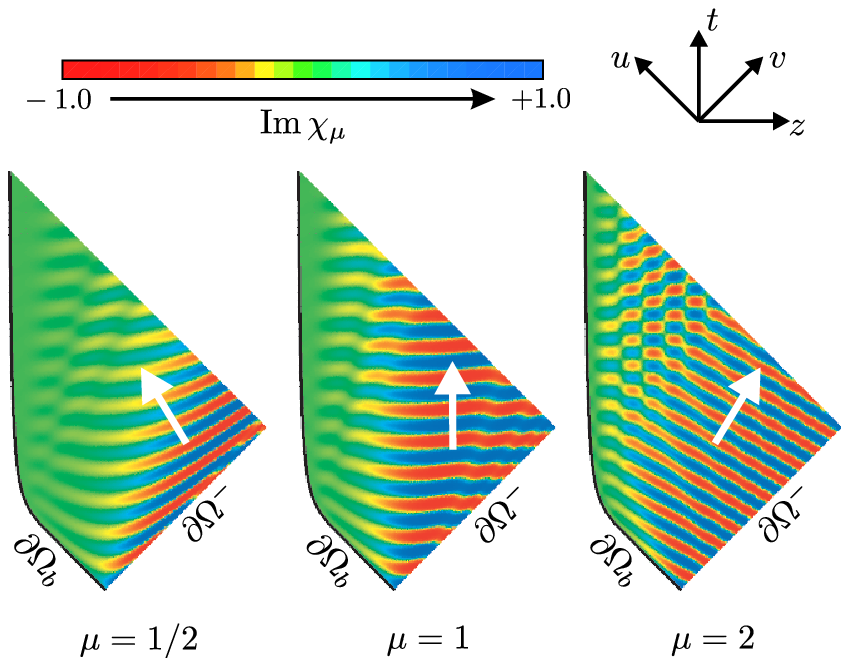}
\caption{(Colour online.) Bulk GW profiles for
$\text{Im}\,\chi_\mu$ initial data.  Here, we have taken
$(\epsilon_*,s_0)=(10,5)$ and $w=1/3$.  The white arrows indicate
the direction of the asymptotic flux derived in
Eq.~(\ref{eq:flux}).}\label{fig:bulk chi compare}
\end{figure}

In Fig.~\ref{fig:bulk chi compare}, we show the bulk GW profile
associated with the imaginary part of $\chi_\mu$ for several
different choices of $\mu$.  In all cases, we see that the GW
profile far from the brane appears to be that of a plane wave, in
agreement with the discussion of the last subsection.  A white
arrow indicating the expected direction of propagation of these
plane waves (cf.~\ref{eq:flux}) is superimposed on each plot; and
we see that there is reasonable agreement between our $\chi_\mu$
approximation (\ref{eq:approximate chi}) and actual simulations
asymptotically.  Intriguingly, we see that as $\mu$ is increased,
more of the initial data seems to `reach' the brane.  Stated in
another way:  When $\mu$ is small, the initial data loses its
coherence when propagating from $\di\Omega^-$ to $\di\Omega_b$.

\begin{figure}
\includegraphics{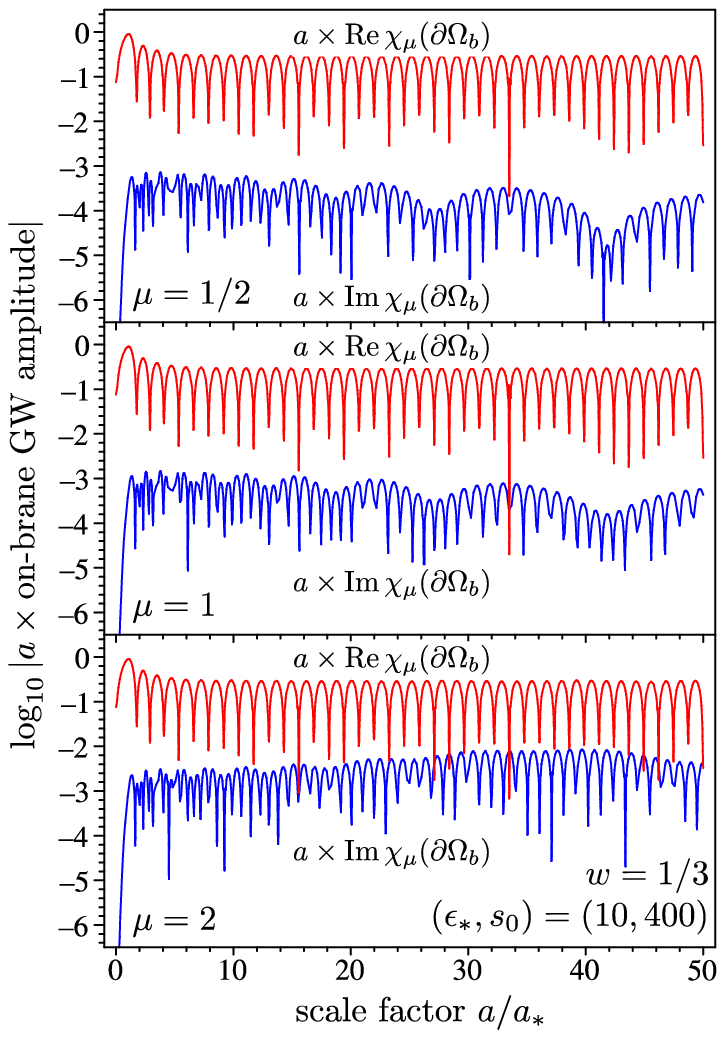}
\caption{(Colour online.) On-brane waveforms for $\chi_\mu$
initial data as $\mu$ is increased and other parameters are kept
constant. One can see that $\text{Re}\,\chi_{\mu}$ is relatively
insensitive to $\mu$, while the opposite is true for
$\text{Im}\,\chi_{\mu}$. Generally speaking, the amplitude of the
imaginary part is less than the real part by several orders of
magnitude.}\label{fig:monochromatic compare}
\end{figure}

In Fig.~\ref{fig:monochromatic compare}, we plot the on-brane
waveforms for cases similar to those shown in Fig.~\ref{fig:bulk chi
compare}. Here, we see that $\text{Im}\,\chi_\mu$ is smaller than
$\text{Re}\,\chi_\mu$ by several orders of magnitude, which implies
that
\begin{multline}
    \text{Re}\,\chi_\mu(\di\Omega_b) \gg \text{Im}\,\chi_\mu(\di\Omega_b)
    \,\, \Rightarrow \\ \chi_\mu(\Omega_b) \approx
    \text{Re}\,\chi_\mu(\di\Omega_b).
\end{multline}
That is, at first approximation, $\chi_\mu (\di\Omega_b)$ is
independent of $\text{Im}\,\chi_\mu$.  Furthermore, the real part of
$\chi_\mu(\di\Omega_b)$ shows virtually no variation as $\mu$ is
increased. This leads us to hypothesize that the brane waveforms are
principally determined by value of the initial data \emph{on the
brane}: The very weak dependence of $\chi_\mu(\di\Omega_b)$ on
$\text{Im}\,\chi_\mu$ implies that the brane signal is insensitive
to data with no initial amplitude on the brane, while the
insensitivity of $\text{Re}\,\chi_\mu(\di\Omega_b)$ to $\mu$ implies
that the brane signal does not overtly care about the details of
initial profile in the bulk.

We now want to describe how the late time amplitude of the real and
imaginary parts of $\chi_\mu$ depend on $s_0$ and $\mu$. But there
is small problem: It is apparent from Fig.~\ref{fig:monochromatic
compare} that the late time brane behaviour of $\text{Im}\,\chi_\mu$
is much more complicated than that of $\text{Re}\,\chi_\mu$. Hence,
it is more difficult to obtain the asymptotic amplitude of the
imaginary parts without running simulations for a very long time,
which is computationally expensive.  However, if we are just
interested in a rough characterization of the late time amplitude,
we can define the expansion-normalized average power as
\begin{equation}
    \langle a^2 \text{Re}^2\,\chi_\mu \rangle_b =
    \frac{1}{a_2-a_1} \int_{a_1}^{a_2} da \left[ \frac{a}{a_*}
    \text{Re}\,\chi_\mu(\di\Omega_b) \right]^2,
\end{equation}
with a similar expression for $\text{Im}\,\chi_\mu$.  For our
calculations, we select the lower integration limit to correspond
to an epoch well into the low-energy regime:
\begin{equation}
    1 \gg \delta = \frac{\rho_1}{\lambda} = \epsilon_* \left(
    \frac{a_*}{a_1} \right)^4.
\end{equation}
On the other hand, $a_2$ is selected to be as large as is
computationally feasible.  Roughly speaking, $\sqrt{2 \times
\langle\text{power}\rangle}$ gives the characteristic amplitudes
seen in Fig.~\ref{fig:monochromatic compare}.

\begin{figure}
\begin{center}
\subfigure[\,\,Average power as a function of initial wavelength perpendicular to the brane]%
{\includegraphics{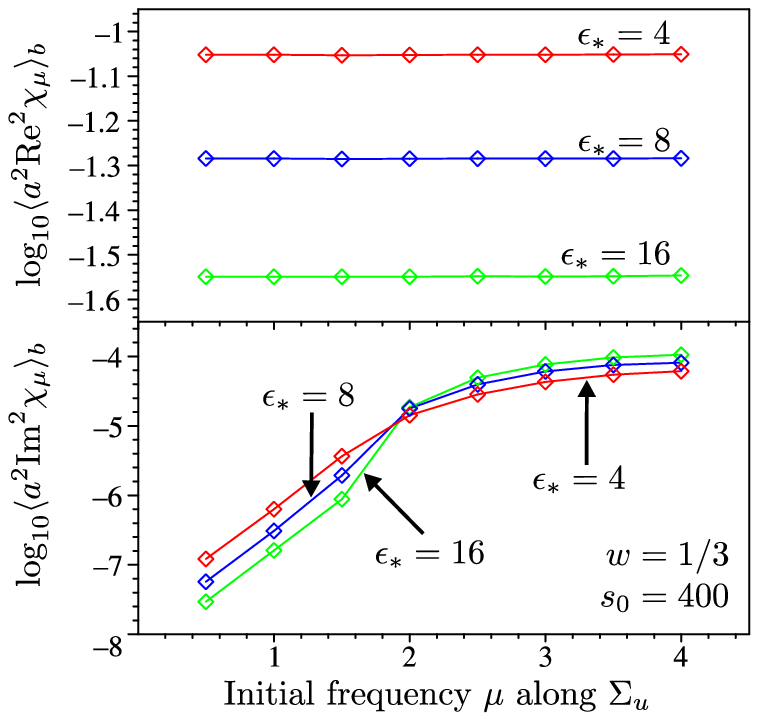}\vspace{5mm}%
\label{fig:average power mu}}%
\\ %
\subfigure[\,\,Average power as a function of initial wavelength parallel to the brane]%
{\includegraphics{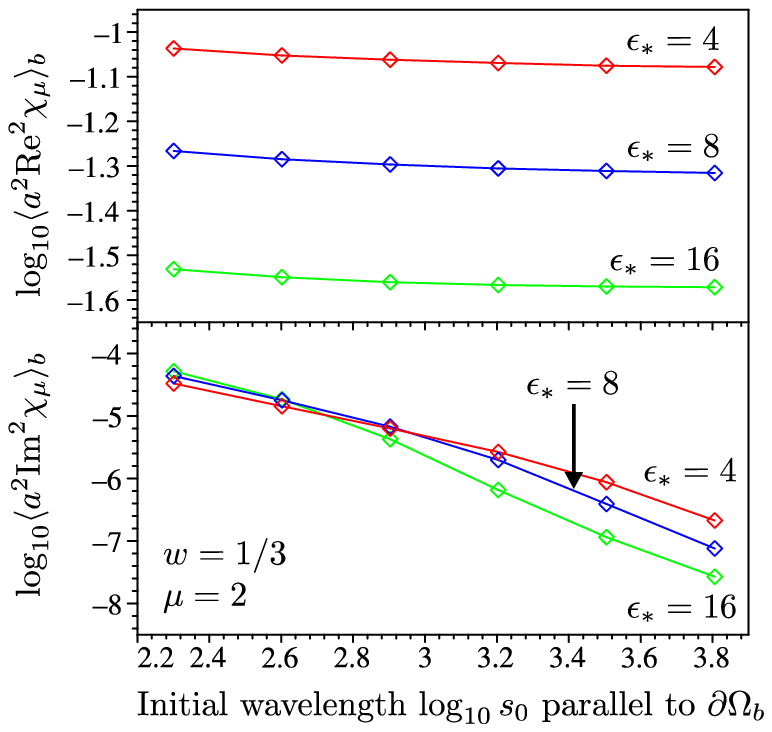}%
\label{fig:average power s_0}}\vspace{5mm}%
\caption{(Colour online.)
Expansion-normalized average power of the real and imaginary parts
of $\chi_\mu$ as functions of parameters. To calculate the average,
we have selected $\delta = 10^{-3}$ and $a_2 = 50a_*$.  The average
power at late times is not very sensitive to these
choices.}\label{fig:average power}
\end{center}
\end{figure}

Using the average power statistic as defined above, we study the
dependence of the late time waveforms on parameters in
Fig.~\ref{fig:average power}.  In both panels, we see that
$\text{Re}\,\chi_\mu$ does not show much variation with either $\mu$
or $s_0$ for several different values of $\epsilon_*$.  However, the
imaginary parts of the basis functions change by several orders of
magnitude as the parameters are varied.  The principal trends are
for $\langle a^2 \text{Im}^2\chi_\mu \rangle$ to increase with $\mu$
and decrease with $s_0$.

The physical inferences we can draw from the numeric work of this
subsections is as follows:  The late time behavior of perturbations
on the brane is largely fixed by the value of initial data at the
brane for a large range of parameters and choices of $h_i(v)$. The
degree to which $h_b$ is oblivious to the bulk initial data profile
increases as the initial data surface is pushed further and further
into the past (cf.~Fig.~\ref{fig:domain}).  Conversely, we find that
these conclusions are mitigated if the Fourier spectrum of the
initial data along $\di\Omega^-$ involves high frequencies.  In such
cases, the bulk initial data finds it `easier' to reach the brane
directly.

The main observational inference of our simulations is that, for
$\mu$ less than some cutoff $\mu_c$, the late time behaviour of
the basis functions independent of $\mu$ and given by
\begin{equation}\label{eq:approximate asymptotic}
    \chi_\mu(\di\Omega_b) \xrightarrow[\infty]{\eta} \frac{\mathcal{C}_b}{a}
    \cos(k\eta + \varsigma), \quad \text{for all }\mu \lesssim \mu_c.
\end{equation}
In general $\mu_c$ will be a function of $s_0$, and we expect that
$\mu_c \rightarrow \infty$ for $s_0 \rightarrow \infty$.  The exact
definition of $\mu_c$ will depend on how rigorously one wants to
enforce the approximate asymptotic form (\ref{eq:approximate
asymptotic}).  Now, if the amplitude components of a particular
initial data profile satisfy
\begin{equation}\label{eq:independence condition}
    \mathcal{A}(\mu) \approx 0 \text{ for } \mu \gtrsim \mu_c,
\end{equation}
the late time waveform follows directly from (\ref{eq:h
decomposition}):
\begin{equation}
    h_b \xrightarrow[\infty]{\eta} h_i(v_i) \times \frac{\mathcal{C}_b}{a}
    \cos(k\eta + \varsigma),
\end{equation}
where $h_i(v_i)$ is just the initial data at the brane position. By
definition, the reference wave introduced in Sec.~\ref{sec:HKT vs
IN} is also linear in $h_i(v_i)$, so the 5D/4D amplitude ratio
$\mathcal{R}$ --- and hence $\Omega_\text{GW}$ --- will be
independent of the details of the initial data profile, provided
that the condition (\ref{eq:independence condition}) is met. In
other words, for finite $s_0$ we expect the late time gravitational
wave spectrum to be independent of the detailed shape of the initial
data profile, provided that that profile does not involve high
frequency features.  From this, it follows that the prediction of a
flat GW spectrum derived from the \citetalias{Hiramatsu:2004aa} and
\citetalias{Ichiki:2004sx} formulations will generalize to more
generic initial data.

\subsection{Analytic results}

The inferences of the last subsection are just that: informed
intuition about the behaviour of initial data based on the
experience gleaned from a finite number of simulations.  We can put
them on somewhat surer footing by doing some approximate analytic
calculations, which is the purpose of this subsection.

\begin{figure}
\includegraphics{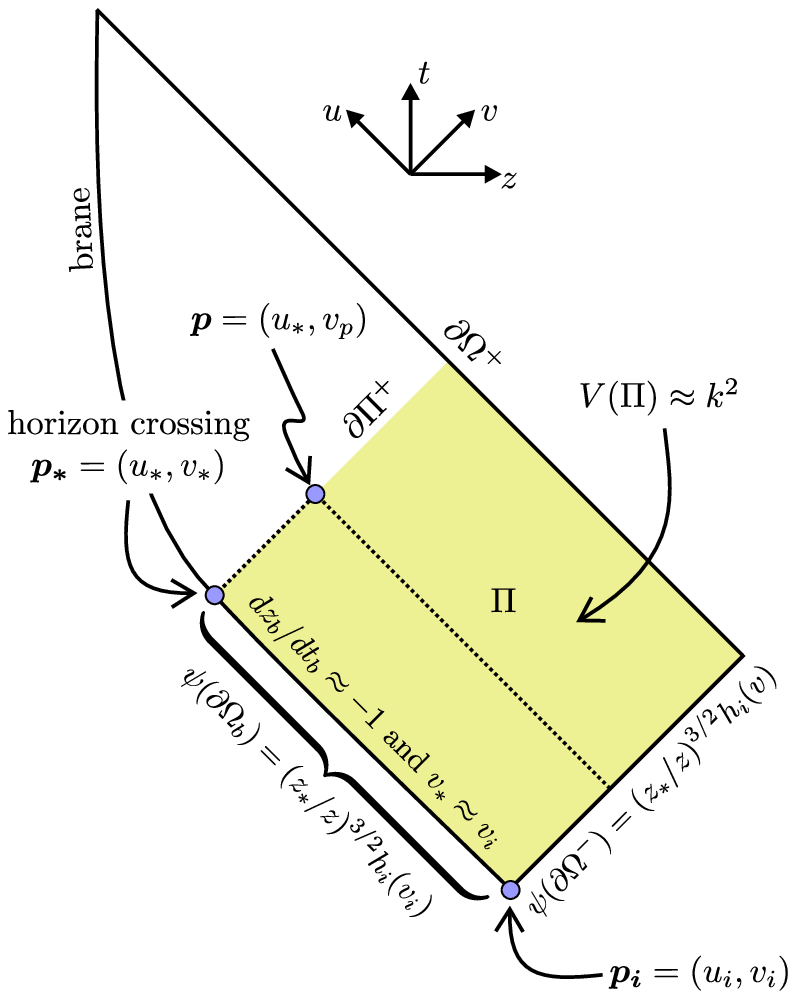}
\caption{(Colour online.) Spacetime geometry used to rationalize the
behaviour seen in the numeric simulations of Sec.~\ref{sec:numerical
results} using an approximate retarded Green's
function.}\label{fig:greens function}
\end{figure}

Let us analyze the behaviour of the wavefunction $\psi$ in the
high-energy epoch of the cosmological evolution.  To be more
precise, we limit our attention to modes with $\epsilon_* \gg
\epsilon_c \approx 0.41$; i.e., modes that have $H_* \ell \gg 1$.
This implies that the slope of the bulk brane trajectory when $a =
a_*$ is
\begin{equation}
    \left( \frac{dz_b}{dt_b} \right)_{a=a_*} =
    -\frac{H_*\ell}{\sqrt{1+(H_*\ell)^2}} \approx -1.
\end{equation}
Hence, the brane trajectory is very nearly null for $a \lesssim
a_*$; here, we will boldly assume that it is exactly null. The
computational domain $\Omega$ for this situation is illustrated in
Fig.~\ref{fig:greens function}. We label the brane's position at
horizon re-entry as $\bm{p_*} = (u_*,v_*)$, and define $\Pi$ to be
the portion of $\Omega$ located to the past of $u = u_*$.  Our goal
is to approximate the field on the future boundary $\di\Pi^+$ of
$\Pi$ given initial data on $\di\Omega^-$.

Note that since we are assuming that $\epsilon_*$ is large, then
$kz_* = \sqrt{\epsilon_*(2+\epsilon_*)}$ is also large; i.e., $k
\gg 1/z_*$. This means that the potential in the bulk wave
equation (\ref{eq:wave eqns}) satisfies
\begin{equation}
    V(z) = k^2 + \frac{15}{4z^2} \approx k^2, \quad \text{for } z
    \ge z_*.
\end{equation}
With this, we can immediately write down an approximate retarded
Green's function for the bulk wave equation with $(\bm{p},\bm{p'})
\in \Pi$:
\begin{multline}
    G(\bm{p};\bm{p'}) =
    \frac{1}{2} J_0(k\lambda) \theta(u-u') \theta(v-v'), \\
    (-\bm{D}^2 + k^2)\,G(\bm{p};\bm{p'}) = \delta(\bm{p}-\bm{p'}).
\end{multline}
Here, we have defined $\lambda = \sqrt{(u-u')(v-v')}$ which is the
proper time interval between the source point $\bm{p'}$ and the
field point $\bm{p}$. In terms of this Green's function, we can
express the value of $\psi$ at any $\bm{p} \in \Pi$ as:
\begin{equation}\label{eq:green's evolution}
    \psi(\bm{p}) \approx \int\limits_{\di\Pi} \bm{n \cdot} \left[
    G(\bm{p};\bm{p'}) \bm{D'} \psi(\bm{p'}) - \psi(\bm{p'})
    \bm{D'} G(\bm{p};\bm{p'}) \right].
\end{equation}
Here, the integration is over $\bm{p'}$, $\bm{n}$ is the outward
pointing normal, and $\bm{D'}$ indicates the derivative with respect
to $\bm{p'}$; i.e., differentiation with respect to the primed
coordinates.  The approximation sign comes from the fact that $G$ is
not the `true' Green's function.

Let us now push $\bm{p}$ to the future boundary $\di\Pi^+$ of
$\Pi$.  Since the Green's function has support when $\bm{p}$ is in
the future of $\bm{p'}$, to calculate the integral
(\ref{eq:green's evolution}) we need only specify the field values
on $\di\Omega^-$ and the portion of the brane $\di\Omega_b$ to the
past of $\di\Pi^+$.  We leave the initial data on $\di\Omega^-$
arbitrary:
\begin{equation}
    \psi(\di\Omega^-) = \left(\frac{z_*}{z} \right)^{3/2} h_i(v).
\end{equation}
However when the brane trajectory is null, the boundary condition
(\ref{eq:RC BC}) reduces to
\begin{equation}
    \left( \frac{\di h}{\di u} \right)_b \approx 0 \,\, \Rightarrow \,\, \psi(\di\Omega_b)
    \approx \left(\frac{z_*}{z} \right)^{3/2} h_i(v_i);
\end{equation}
i.e., $h$ is constant on the brane before horizon crossing, which
is entirely consistent with our numeric simulations.  Using this
boundary data, simplifying the integral (\ref{eq:green's
evolution}) is straightforward, but tedious.  The result is:
\begin{multline}
    \psi(\bm{p}) \approx (2z_*)^{3/2} \Bigg[ 3  h_i(v_i)
    \int\limits_{u_i}^{u_*} du'
    \frac{J_0(k\lambda_1)}{(v_*-u')^{5/2}} + \\
    \int\limits_{v_i}^{v_p} dv' J_0(k\lambda_2) \underbrace{\frac{\di}{\di v'}
    \frac{h_i(v')}{(v'-u_i)^{3/2}}}_{\mathcal{O}(h_i'/u_i^{3/2})}  -  J_0(k\lambda_3)
    \underbrace{\frac{ h_i(v_i)}{(v_i-u_i)^{3/2}}}_{\mathcal{O}(h_i/u_i^{3/2})}
    \Bigg], \label{eq:psi approximation}
\end{multline}
where we have defined
\begin{gather}
    \nonumber \lambda_1 = \sqrt{(u_*-u)(v_p-v_*)}, \\
    \nonumber \lambda_2 = \sqrt{(u_*-u_i)(v_p - v)}, \\
    \lambda_3 = \sqrt{(u_*-u_i)(v_p - v_i)}.
\end{gather}
To arrive at this expression, we have used integration by parts to
remove derivatives of the Green's function.

The approximation (\ref{eq:psi approximation}) can be used to
justify the trends we have seen above, if we note that both the
initial data hypersurface and horizon crossing epoch are in the
high-energy regime:
\begin{equation}
    s_0 = \frac{H_i a_i}{H_* a_*} \approx
    \sqrt{\frac{\epsilon_*}{2+\epsilon_*}} \left( \frac{v_i -
    u_i}{v_* - u_*} \right)^3 \xrightarrow[-\infty]{u_i} \mathcal{O}(-u_i^3).
\end{equation}
When inserted in (\ref{eq:psi approximation}), this yields
\begin{multline}\label{eq:psi approximation 2}
    \psi(\bm{p}) \approx (2z_*)^{3/2} 3  h_i(v_i)
    \int\limits_{u_i}^{u_*} du'
    \frac{J_0(k\lambda_1)}{(v_*-u')^{5/2}} \\ +
    \int dv\, \mathcal{O}(h_i' s_0^{1/2})
    + \mathcal{O}(h_i s_0^{1/2}).
\end{multline}
If we then hold the initial $h_i(v)$ profile constant, as in
Fig.~\ref{fig:average power s_0}, we see that the last two terms
drop out as $s_0 \rightarrow \infty$. Hence, we have shown that as
the initial data surface goes to the infinite past, the wavefunction
on $\di\Pi^+$ depends only on the value of the initial data on the
brane $h_i(v_i)$ --- and not on the initial data in the bulk $h_i(v
> v_i)$.  Since the evolution of the GWs to the future of $\di\Pi^+$
depends only on the field value on $\di\Pi^+$, we can conclude that
the entire late time brane waveform is determined by $h_i(v_i)$ in
the limit $s_0 \rightarrow \infty$.  This explains the behaviour
seen in Fig.~\ref{fig:average power s_0}.

Now, consider the situation if we hold the position of the initial
data hypersurface constant and vary the initial data profile, as in
Fig.~\ref{fig:average power mu}.  Let us also assume that $s_0$ is
large and that $h_i(v) = \mathcal{O}(1)$.  Then, the third term on
the righthand side of (\ref{eq:psi approximation 2}) is not
important relative to the first.  However, the second term can be
arbitrarily large if we allow for initial data with large gradients
$h'_i(v)$. In particular if we have $h_i(v) = e^{i\mu k(v+u_i)/2}$,
which is the initial data that generates $\chi_\mu$, we see that the
second term grows with $\mu$.  Hence, we have shown that if the
initial data on $\di\Omega^-$ involves a significant high frequency
component, then the late time brane waveform will be sensitive to
the precise form of $h_i(v)$.  This rationalizes the behaviour seen
in Fig.~\ref{fig:average power mu}.

\section{Conclusions}\label{sec:conclusions}

In this paper, we have developed a new numeric algorithm (from first
principles) to deal with the problem of solving $(1+1)$-dimensional
wave equations in the presence of a moving boundary.  Our technique,
which is based on characteristic integration methods from black hole
perturbation theory, has demonstrated itself to be both reliable and
accurate by reproducing previously known analytic and numeric
results; and it is the principle result of this work.

We have applied our formalism to the cosmological problem of the
evolution of GWs in the Randall-Sundrum one-brane scenario.  One
can find at least three different prescriptions in the literature
\cite{Ichiki:2004sx,Kobayashi:2005dd,Hiramatsu:2004aa} for how to
specify post-inflationary bulk initial conditions in such models,
which yield at least two contradictory predictions for the
spectral tilt of the stochastic GW background generated by brane
inflation.  Until now, it was unclear if the discrepancy between
various results was due to different choices of initial data or
the numeric scheme used to evolve GWs through the high-energy
radiation era after inflation.  Using our code, we have
investigated the initial conditions proposed by
\citetalias{Hiramatsu:2004aa} and \citetalias{Ichiki:2004sx}, and
we find no discrepancy between the late time GW spectrum generated
by either prescription, provided that the initial data
hypersurface is set far enough into the past.  Observationally,
this means that both initial conditions lead to a flat GW spectrum
$\Omega_\text{GW} \propto (f/f_c)^0$ at frequencies higher that a
threshold $f_c$ set by the curvature scale of the bulk.  This is
also in agreement with the results of
\citetalias{Kobayashi:2005dd}, who study the quantum (i.e.~not
classical) evolution of the GW wavefunction.

We have also considered more general initial data by introducing a
practical basis in which to decompose general solutions of the
wave equation.  Numerically, we find that the late time brane
waveform is not very sensitive to the detailed initial data
profile if we start our simulations at sufficiently high energies.
However, this approximation breaks down if the Fourier transform
of the initial data involves very high frequencies.  We have used
an approximate Green's function analysis to analytically
rationalize these results and demonstrate how they apply to
\emph{any} initial data; not just the choices we modelled
explicitly.\footnote{Note that this is in general agreement with
the recent (numerical) work of \citet{Kobayashi:2006pe} using a
quantum formalism.  But note that our result differs in that we
consider completely arbitrary initial conditions, while
\citeauthor{Kobayashi:2006pe} considers initial conditions from
pure de Sitter inflation.}

\begin{figure}
\includegraphics{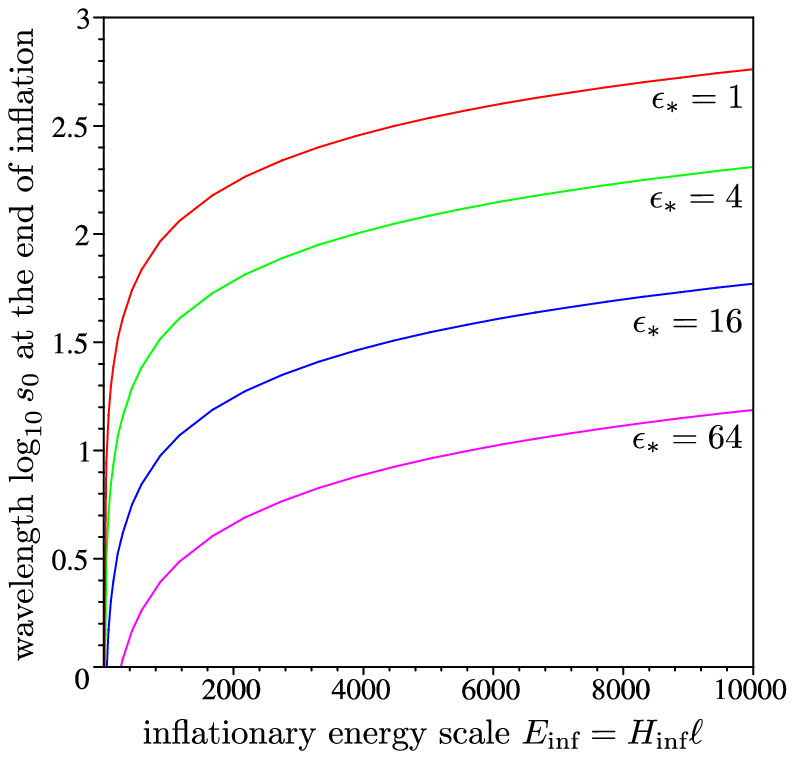}
\caption{(Colour online.) Wavelength of perturbations at the end
of inflation as a function of the inflationary energy
scale.}\label{fig:inflation energy scale}
\end{figure}
So, is the high-frequency stochastic GW spectrum predicted from
this class of braneworld models really flat?  The answer seems to
depend rather intimately on the energy scale of inflation.
Throughout this paper, we have treated $s_0$ as a free parameter,
and our results on the insensitivity to initial data depend on the
limit $s_0 \rightarrow \infty$.  But really, we should fix the
energy scale of inflation $E_\text{inf} \equiv H_\text{inf} \ell$
and synchronize $\di\Omega^-$ with the beginning of the
high-energy radiation era.  This means that $s_0$ is actually a
function of $\epsilon_*$ and $E_\text{inf}$:
\begin{equation}
    s_0(\epsilon_*,E_\text{inf}) =
    \frac{E_\text{inf}^{3/4}}{\sqrt{\epsilon_*(2+\epsilon_*)^2}} \left[
    \frac{E_\text{inf}}{\sqrt{1+E_\text{inf}^2} - 1}
    \right]^{1/4}.
\end{equation}
For reference, we have plotted $s_0$ as a function of
$E_\text{inf}$ in Fig.~\ref{fig:inflation energy scale}.  As can
be seen from this plot, if we consider moderate inflationary
energy scales $E_\text{inf} \lesssim 10^3$, it is possible to have
$s_0 \lesssim 10^2$ for reasonable values of $\epsilon_*$.  The
results of Sec.~\ref{sec:generic initial data} imply that such
values of $s_0$ imply the dependence of the late time waveforms on
initial data is weak, but non-trivial.  This suggests to us that
the greatest chance of obtaining deviations from the flat spectrum
lies in models with small inflationary energy scales.  In such
scenarios, it is possible for the brane signal to carry a
signature of $h(\di\Omega^-)$, which in turn depends on the
details of the inflationary model.  On the other hand, if the
inflationary energy scale is high, the brane signal will only
depend on the value of the perturbation on the brane at the end of
inflation. Investigating the detailed gravitational wave spectrum
generated by moderately low-energy brane inflation is an
interesting avenue for future work.

\begin{acknowledgments}
I would like to thank Takashi Hiramatsu, Kazuya Koyama, and Atsushi
Taruya for sharing the results of their numerical calculations and
useful discussions.  I would also like to thank Chris Clarkson,
Tsutomu Kobayashi and Roy Maartens for comments.  I am supported by
PPARC postdoctoral fellowship PP/C001079/1.
\end{acknowledgments}

\appendix

\section{Characterizing the gravity wave background}\label{sec:GW
background}

In this appendix, we review how to map the results of our
calculations into observable predictions concerning the spectrum
of the cosmological GW background today.  The treatment is very
much the same as in Refs.~\cite{Hiramatsu:2004aa,Ichiki:2004sx}.

On the brane, the complete GW perturbation is written as
\begin{equation}
h_{ij}(\tau,\mathbf{x}) = \frac{a^2}{(2\pi M_5)^3}
\sum_{A=+,\times} \int d^3k \, h_b(\tau;\mathbf{k},A)
e^{i\mathbf{k} \cdot \mathbf{x}}
\varepsilon^{(A)}_{ij}(\mathbf{\hat{k}}),
\end{equation}
Here, $\tau$ is the cosmic time; i.e., $ds^2 = -d\tau^2 + a^2
d\mathbf{x}^2$ on the brane.  The energy density associated with
this perturbation is
\begin{equation}
    \rho_\text{GW} = \frac{1}{4\kappa_4^2} \bigg\langle \frac{d
    h^*_{ij}}{d\tau} \frac{dh^{ij}}{d\tau} \bigg\rangle,
\end{equation}
where the angular brackets indicate an average over some region of
spacetime.  To preform the spatial average, we make the standard
assumption that the background is unpolarized, stationary, and
isotropic.  This means we can trade the operation of spatial
averaging over $\mathbf{x}$ for ensemble averaging at
$\mathbf{x}=0$, and the Fourier amplitudes obey:
\begin{multline}
    \langle h_b(\tau;\mathbf{k},A) h_b^*(\tau;\mathbf{k}',A')
    \rangle = \\
    (2\pi M_5)^3 \delta_{AA'} \delta ( \mathbf{k} - \mathbf{k}') |h_b(\tau;
    \mathbf{k},A)|^2.
\end{multline}
The product of polarization tensors can then be reduced via
\begin{equation}
    \varepsilon^{(A)}_{ij}(\mathbf{\hat{k}}) \, \varepsilon^{(A)ij}(\mathbf{\hat{k}}) = 2.
\end{equation}
Temporal averaging in the late universe can be done by noting that
in the late universe, our numeric results give that $h_b$ is a
superposition of the $e^{\pm ik\eta}/a$ `zero-mode' functions.
Neglecting scale factor derivatives and using $\eta \approx
\tau/a$ yields
\begin{equation}
    \bigg\langle \left| \frac{d h_b (\tau;\mathbf{k},A)}{d\tau} \right|^2 \bigg \rangle_\tau =
    \frac{k^2\mathcal{C}_b^2(k)}{2a^4},
\end{equation}
where $\mathcal{C}_b(k)$ is the expansion-normalized
characteristic amplitude (cf.~Eq.~\ref{eq:4D asymptotic}). To
relate $\mathcal{C}_b(k)$ to the primordial GW spectrum, we make
use of the reference wave $h_\text{ref}$ discussed in
Sec.~\ref{sec:HKT vs IN}.  Our numeric simulations can be used to
find
\begin{equation}
    \mathcal{R} =
    \frac{\mathcal{C}_b} {\mathcal{C}_\text{ref}};
\end{equation}
i.e., the ratio of the characteristic amplitudes in the low-energy
regime.  This is useful because the evolution of the reference wave
through the high-energy radiation epoch is extremely simple:
Essentially, it remains constant until $a = a_*$, then its amplitude
decays as $1/a$.  Hence, we can write
\begin{equation}
    \mathcal{C}^2_\text{ref}(k) \approx a^2_* \mathcal{C}^2_i(k) =
    (k/H_*)^2 \mathcal{C}^2_i(k).
\end{equation}
Here, $\mathcal{C}_i^2(k)$ is the squared amplitude of $h_b$, set
after inflation.  (Recall that, by definition, $h_b$ and
$h_\text{ref}$ are identical before horizon re-entry, which means
the share the same initial power spectrum.)  We can conveniently
re-express this in terms of
\begin{equation}
    \delta_T^2 = \frac{8\pi k^3}{(2\pi M_5)^3} \mathcal{C}^2_i,
\end{equation}
which for the inflationary primordial spectrum discussed in
Sec.~\ref{sec:initial conditions} reduces to
\begin{equation}\label{eq:delta_T from inflation}
    \delta_T^2 = 2\kappa_4^2 C^2(H_\text{inf}\ell) \left( \frac{H_\text{inf}}{2\pi}
    \right)^2,
\end{equation}
where $H_\text{inf}$ is the Hubble parameter during inflation, and
we have made use of $M_5^3 \kappa_4^2 \ell = 1$. Putting all these
results together yields
\begin{equation}
    \rho_\text{GW} = \frac{
    1}{8\kappa_4^2 a^4} \int d\ln k \, k^2
    \left( \frac{k}{H_*} \right)^2 \delta_T^2 \mathcal{R}^2.
\end{equation}
For comparison to actual experiments, it is convenient to re-express
this in terms of the frequency $f = k/2\pi a$ observed today and
introduce the spectral density
\begin{equation}
    \Omega_\text{GW} = \frac{1}{\rho_c} \frac{d\rho_\text{GW}}{d\ln
    f} = \frac{2\pi^4 f^4 \delta_T^2(f) \mathcal{R}^2(f)}{3 H_0^2 H_*^2(f)}.
\end{equation}
where $\rho_c = 3H_0^2/\kappa_4^2$ is the critical density.

To progress further, we need to know $H_* = H_*(f)$; that is, for
a given mode with frequency $f$, we need to know the Hubble length
when it re-entered the horizon.  It is useful to introduced a
\emph{critical frequency}, which corresponds to the mode
re-entering the horizon when $H\ell = 1$.  As discussed in
Sec.~\ref{sec:RS problem}, this mode has $\epsilon_* = \epsilon_c
= \sqrt{2} - 1$.  We can measure all other frequencies as a
multiples of the critical frequency via
\begin{equation}\label{eq:f/f_c ratio}
    \frac{f}{f_c} = \frac{H_* a_*}{H_c a_c} = \left[ \epsilon_* (\sqrt{2}-1)(2 + \epsilon_*)^2 \right]^{1/4}.
\end{equation}
Here, we have used that $\epsilon_c a_c^4 = \epsilon_* a_*^4$. This
is a cubic equation in $\epsilon_*$ that can be analytically
inverted to give $\epsilon_* = \epsilon_*(f/f_c)$, which then yields
$H_*\ell$ as a function of $f/f_c$.  However, the general formula is
complicated and not particularly enlightening. More interesting are
the limits:
\begin{equation}
    H_*\ell \approx
    \begin{cases}
        \sqrt{ 1+ \frac{1}{\sqrt{2}} } \left( \frac{f}{f_c}
        \right)^{2}, & f \lesssim f_c.
        \\ (\sqrt{2}+1)^{1/3} \left( \frac{f}{f_c} \right)^{4/3}, & f \gtrsim f_c.
    \end{cases}
\end{equation}
This yields that
\begin{equation}\label{eq:spectral density}
    \Omega_\text{GW} \approx \frac{f_c^4 \ell^2 }{H_0^2} \delta_T^2(f) \mathcal{R}^2(f)
    \begin{cases}
        54.9, & f \lesssim f_c.
        \\ 36.4 (f/f_c)^{4/3}, & f \gtrsim f_c.
    \end{cases}
\end{equation}
Hence, in order to understand the the frequency dependence of
$\Omega_\text{GW}$, we need to know $\mathcal{R} = \mathcal{R}(f)$
from numeric simulations.

Finally, we need to specify the actual value of $f_c$.  We make use
of the fact that the universe expands adiabatically during radiation
and matter domination.  Therefore, conservation of entropy yields
\begin{equation}
    g_S(T_c) a_c^3 T_c^3 = g_S(T_0) a_0^3 T_0^3.
\end{equation}
Here, $T_c$ and $T_0$ indicate the temperature at the critical epoch
$H_c \ell = 1$ and today respectively, and $g_S$ measures the
effective number of degrees of freedom in the matter sector as a
function of temperature.  We can relate the temperature of radiation
at the critical epoch to its density via
\begin{equation}
    \rho_c = \lambda \epsilon_c = \frac{6
    \epsilon_c}{\kappa_4^2 \ell^2} = \frac{g_c \pi^2
    T_c^4}{30},
\end{equation}
where we have written $g_c = g_S(T_c)$.  Then, it follows that
\begin{multline}
    f_c = \frac{k}{2\pi a_0} = \frac{H_c a_c}{2\pi a_0} =
    \frac{1}{2\pi\ell} \frac{g_0^{1/3} T_0}{g_c^{1/3} T_c} = \\
    \frac{1}{2(180\pi^2\epsilon_c)^{1/4}}
    \frac{g_0^{1/3}}{g_c^{1/12}}
    \left( \frac{\kappa_4}{\ell} \right)^{1/2} T_0.
\end{multline}
To get a numerical answer, we can take $T_0 = 2.75$ K, $g_0 = 3.91$
\cite{Kolb:1990}, and $\epsilon_c = \sqrt{2} - 1$.  Then,
\begin{equation}
    f_c = 3.3 \times 10^{-5} \left( \frac{\text{0.1 mm}}{\ell} \right)^{1/2}
    \left( \frac{100}{g_c} \right)^{1/12} \text{Hz}.
\end{equation}

\bibliography{GW_moving_brane}

\end{document}